\documentclass[12pt]{iopart}

\usepackage{iopams}  
\begin{document}

\title[]{Spacetime Quanta? : The Discrete Spectrum of a Quantum Spacetime Four-Volume Operator in Unimodular Loop Quantum Cosmology}

\author{J. Bunao}

\address{Department of Physics, Ateneo de Manila University, Katipunan Ave., Quezon City, 1108 Philippines}
\ead{jbunao@ateneo.edu}
\vspace{10pt}
\begin{indented}
\item[]November 2016
\end{indented}

\begin{abstract}
This study considers the operator $\hat{T}$ corresponding to the classical spacetime four-volume $\tilde{T}$ (on-shell) of a finite patch of spacetime in the context of Unimodular Loop Quantum Cosmology for the homogeneous and isotropic model with flat spatial sections and without matter sources. Since the spacetime four-volume is canonically conjugate to the cosmological "constant", the operator $\hat{T}$ is constructed by solving its canonical commutation relation with $\hat{\Lambda}$ - the operator corresponding to the classical cosmological constant on-shell $\tilde{\Lambda}$. This conjugacy, along with the action of $\hat{T}$ on definite volume states reducing to $\tilde{T}$, allows us to interpret that $\hat{T}$ is indeed a quantum spacetime four-volume operator. The discrete spectrum of $\hat{T}$ is calculated by considering the set of all $\tau$'s where the eigenvalue equation has a solution $\Phi_{\tau}$ in the domain of $\hat{T}$. It turns out that, upon assigning the maximal domain $D(\hat{T})$ to $\hat{T}$, we have $\Phi_{\tau}\in D(\hat{T})$ for all $\tau\in\mathbb{C}$ so that the spectrum of $\hat{T}$ is purely discrete and is the entire complex plane. A family of operators $\hat{T}^{(b_0,\phi_0)}$ was also considered as possible self-adjoint versions of $\hat{T}$. They represent the restrictions of $\hat{T}$ on their respective domains $D(\hat{T}^{(b_0,\phi_0)})$ which are just the maximal domain with additional quasi-periodic conditions. Their possible self-adjointness is motivated by their discrete spectra only containing real and discrete numbers $\tau_m$ for $m=0,\pm1,\pm2,...$.

\end{abstract}

%
%
%
%
%

\section{Introduction}
The construction of a quantum theory of gravity aims to unify the empirically successful theories of General Relativity (GR) and Quantum Field Theory (QFT). Since according to GR, spacetime represents the gravitational field and is a dynamical quantity much like any other field, and according to QFT, any dynamical field is represented by discrete quanta that can be in a superposition of states, then there should be quanta of spacetime \cite{book1, book2}, like "atoms" of spacetime, subject to the fuzziness of quantum superposition and where the smooth classical spacetime emerges at large scales. Perhaps then, one can consider a patch of spacetime (corresponding to an "atom"), determine its quantum dynamics (analogous to one-particle dynamics), and determine the rules to combine the patches into a many-patch system (analogous to interacting many-body systems) \cite{bojrev1}. Loop Quantum Gravity (LQG) is one such approach where the quantum discreteness of spacetime is derived rather than postulated \cite{book3}. This is apparent in the discrete eigenvalues of the area and volume operators \cite{art1} suggesting that there is a minimum possible area and volume. That is, space is discrete according to LQG. Group Field Theories (GFTs), which are related to Spin Foam Models (SFMs) and can be considered as second quantized formulations of LQG, consider the microscopic interactions of these atoms so that macroscopic spacetime emerges as a condensate \cite{gft1, gft2}. Loop Quantum Cosmology (LQC) is an application of the loop/polymer quantization of LQG on symmetry reduced cosmological models \cite{lqcrev1, lqcrev2, lqcrev3, lqcrev4, lqcrev5, lqcart1,lqcart2, lqcart3, lqcart4, lqcart5, lqcart6}. Note that this may not be the same as doing a symmetry reduction from the full quantum theory. The loop quantization of LQG produces quantum representations inequivalent to those used in a Wheeler-DeWitt (WDW) quantization - a quantization procedure akin to usual quantum mechanics. One motivation for studying LQC is that the simplified models can provide a testing ground for some aspects and ideas from the full theory of LQG. Some of these models are called minisuperspace models - reduced theories with no remaining field-theory degrees of freedom. These theories include the homogeneous and isotropic Friedmann-Lemaitre-Robertson-Walker (FLRW) models for the spatially flat ($k=0$) \cite{lqcbb1, lqcbb2, lqcbb3}, positively curved ($k=1$) \cite{lqckp}, negatively curved ($k=-1$) \cite{lqckn}, and with a non-vanishing cosmological constant \cite{lqccosm} cases. As a next step on extending to more general cases, anisotropic cosmologies such as the Bianchi I \cite{lqcbi1, lqcbi12, lqcbi13, lqcbi14, lqcbi15}, II \cite{lqcbi2}, and IX \cite{lqcbi9} models were also considered. Further still, inhomogeneous midisuperspace models, field theories with few enough Killing vectors to ensure local remaining degrees of freedom, are studied in the context of LQC. A specific example is the polarized Gowdy $T^3$ model \cite{lqcgowdy} corresponding to gravitational waves propagating in a closed expanding universe. The quantum nature of black holes has also been considered \cite{bhrev} through the loop quantization of spherically symmetric spacetimes. A full loop quantization was completed through a partial gauge fixing \cite{bhgf1,bhgf2} to eliminate the diffeomorphism constraint and later, this gauge fixing was circumvented \cite{bhwo1,bhwo2}.

A robust result from the analysis of these models in LQC is the avoidance of classical singularities \cite{lqcrev1, lqcbb1, lqcbb2, lqcbb3} such as the Big Bang and Big Crunch singularities being replaced by a Quantum Big Bounce. This avoidance is attributed to the effects of quantum geometry (the discreteness of space) so that there is an effective repulsive force when the universe reaches a certain minimum size and the energy density of matter reaches a maximum value. This departure of LQC dynamics from the classical one near the singularity is not shared by the WDW dynamics such that the classical singularity is not resolved in the WDW theory. The inequivalence of LQC and WDW theory is apparent in the differences in the Hilbert space structure defining the possible normalizable states of the quantum theory, and the Hamiltonian constraint operator defining the quantum dynamics. As an example, the normalizable states in WDW theory are continuous integrals whereas in LQC, they are discrete sums. Moreover, the Hamiltonian in WDW theory is a continuous differential operator while in LQC, it is a discrete difference operator. The prescription in constructing such an LQC Hamiltonian operator is however, not unique. Issues such as the operator ordering and densitization of the Hamiltonian constraint \cite{lqcpres}, and the dynamical refinements of the discreteness scale \cite{lqcrev2, lqcdyref} exist. Nevertheless, the quantum bounce remains a feature for them. The LQC dynamics also admit an effective theory description obtained using coherent state techniques \cite{lqceff} so that an effective Hamiltonian is obtained which modifies the classical equations of motion. Interestingly, the LQC effective theories for the minisuperspace models very accurately capture the underlying quantum dynamics of sharply peaked universe states that become macroscopic at later times. This accuracy is true even near the bounce point \cite{lqceff1}. A few results include a maximum energy density $\rho_{crit} = 3(\kappa\gamma^2\Delta)^{-1}$ (where $\kappa = 8\pi G$, $\gamma\in\mathbb{R}$ is the Barbero-Immirzi parameter, and $\Delta$ is the minimum eigenvalue of the area operator in full LQG \cite{art1,lqcarea}) for at least the $k=0$ FLRW model \cite{lqcrev3}, exotic singularity resolution for all strong and some weak singularities for the flat \cite{lqcesf} and curved \cite{lqcesc} FLRW models, numerical simulations and effective equations of motion incorporating more general quantum states such as widely spread Gaussian and non-Gaussian states \cite{lqcnum1, lqcnum2, lqceffl}, and the effective dynamics of Bianchi IX \cite{lqceffbi9}. Other recent studies in LQC refer to the return of the self-dual variables ($\gamma\rightarrow i$) \cite{lqccav} as a potential answer to the Barbero-Immirzi parameter ambiguity. One motivation is that unlike the self-dual connection, the real valued connection (that is, with $\gamma\in\mathbb{R}$) does not admit an interpretation as a spacetime gauge field \cite{art2}. The hope is that one might get some insight from self-dual LQC in order to construct a self-dual LQG. Since we do not know the kinematical Hilbert space for a self-dual LQG, a possible insight coming from self-dual LQC are in the papers \cite{lqcsd, lqcasd} where for some cosmological models, their corresponding kinematical Hilbert spaces were constructed by choosing a particular inner product to properly impose the reality conditions in the quantum theory. One of the toy models is the three-dimensional Lorentzian LQG where the spectra of the geometric operators, from being discrete in usual LQG, become continuous in the self-dual version \cite{art3}. A question then arises of whether the discreteness of the area and volume spectra would survive in the full four-dimensional self-dual LQG. Another question regarding the discreteness of the spectra of the LQG geometric operators was posed in \cite{art4, art5} where they investigate whether the said discreteness at the kinematic level would survive at the physical level. 

Unimodular Quantum Gravity (UQG) \cite{uqg1,uqg2,uqg3,uqg4} is the attempt to apply quantization rules to the theory of Unimodular Gravity (UG) - a modification of GR where the determinant of the metric was to be held constant. A different classical formulation by Hennaux and Teitelboim has been constructed where the metric can be varied fully and without restriction so that the theory is fully diffeomorphism invariant \cite{HT}. In UG, the cosmological constant $\Lambda$ from usual GR is instead a dynamical variable conjugate to the spacetime four-volume $T$. The spacetime constancy of $\Lambda$ is alternatively enforced by the equations of motion. The solutions of UG are then equivalent to GR with a non-vanishing $\Lambda$. The four-volume $T$ can also be interpreted as a "cosmological time" so that the corresponding dynamics of the quantum theory can refer to this time and the equations become "Schrodinger-like" \cite{time1}. Thus, UQG seeks to address the Frozen Formalism facet of the Problem of Time (POT) in QG \cite{time2}. However, the interpretation of $T$ as the reference time to use still has difficulties as a resolution to the POT \cite{time3}. There are also studies that incorporate the principles of LQG \cite{ulqg1,ulqg2} into UG. Specifically in \cite{ulqc}, the $k=0$ FLRW model was considered and the role of $T$ as time was compared with the usual massless scalar field as the "emergent time". The resulting theory was relatively simpler in the sense that since the Hamiltonian is linear in $\Lambda$, the corresponding time evolution is of first order $\frac{\partial}{\partial T}$ as opposed to the "Klein-Gordon-like" second order evolution arising from the massless scalar field clock.

This study would be in the context of Unimodular Loop Quantum Cosmology (ULQC) as well. Specifically, we will consider the $k=0$ FLRW model and study an operator corresponding to the four-volume $\tilde{T}$ (on-shell), denoted by $\hat{T}$. We will investigate the possible quantum discreteness of a spacetime patch by asking whether $\hat{T}$ has a purely discrete spectrum, in analogy to the discrete spectra of, say, the volume operator of the usual LQC. Also, since the Hamiltonian operator becomes a difference operator as mentioned earlier, perhaps a consequence of the spacetime four-volume discreteness is a discreteness in the time evolution \cite{lqcart5}. That is, $\frac{\partial\psi}{\partial T}$ would be replaced by some $\frac{\Delta\psi}{\Delta T}$ where $\Delta T$ is some minimum four-volume. This mirrors a procedure in usual LQC where the curvature cannot be expressed in terms of holonomies around loops of vanishing area but rather in terms of those around a minimum possible area $\Delta$. This study is structured as follows. In Section \ref{ugav}, we review the theory of UG proposed by Hennaux and Teitelboim and express them in terms of the Ashtekar Variables. In Section \ref{cosm} we apply the theory on the $k=0$ FLRW model and construct an expression for the classical spacetime four-volume on-shell, which we denote as $\tilde{T}$. In Section \ref{hilb}, we review the kinematic Hilbert space $\mathcal{H}^{kin}$ by defining its inner product and a basis for the states in $\mathcal{H}^{kin}$, along with the basic operators acting in $\mathcal{H}^{kin}$. In Section \ref{Top}, we construct a quantum spacetime four-volume operator $\hat{T}$ acting in $\mathcal{H}^{kin}$ with calculation details in \ref{consT}. In Section \ref{disc}, we calculate the discrete spectrum of $\hat{T}$ by solving for the formal solutions $\Phi_{\tau}$ to eigenvalue equation $ (\tau \hat{\mathbb{I}} -\hat{T})\Phi_{\tau} = 0$ and finding the set of all $\tau$'s where $\Phi_{\tau}$ is in the domain of $\hat{T}$ which we set to be its maximal domain $D(\hat{T})$. It turns out that all the $\Phi_{\tau}$'s are normalizable in $\mathcal{H}^{kin}$ so that $\Phi_{\tau}\in D(\hat{T})$ for all $\tau\in\mathbb{C}$. Thus, $\hat{T}$ has the entire complex plane as its purely discrete spectrum. In Section \ref{selfadj}, we consider a family of operators $\hat{T}^{(b_0,\phi_0)}$ as possible self-adjoint versions of $\hat{T}$. The $\hat{T}^{(b_0,\phi_0)}$'s represent the restrictions of $\hat{T}$ on domains with additional quasi-periodic conditions and these domains may be related to the superselection sectors determined by the Hamiltonian constraint of the theory. We motivate their possibility to be self-adjoint by calculating their discrete spectra and seeing that they contain only real and discrete numbers $\tau_m$ for $m=0,\pm1,\pm2,...$. To verify this self-adjointess however, one needs to show that the $\hat{T}^{(b_0,\phi_0)}$'s satisfy certain conditions which will be the topic of future studies. Lastly, in Section \ref{conc}, we conclude.

\section{Classical Unimodular Gravity in terms of the Ashtekar Variables}\label{ugav}
A fully diffeomorphism invariant formulation of unimodular gravity is defined by the Hennaux-Teitelboim (HT) action \cite{HT} 
\begin{equation}\label{htaction}
\mathcal{S} = \frac{1}{2\kappa}\int_{\mathcal{M}} d^4x \left[(R - 2\Lambda)\sqrt{-g} + 2\Lambda \partial_{\mu}\tau^{\mu} \right]
\end{equation}
over a manifold $\mathcal{M}$ which, for simplicity, we consider has a vanishing stress energy tensor. Note that $\kappa = 8\pi G$, $g$ is the determinant of the metric $g_{\mu\nu}$, $R$ is the Ricci Curvature Scalar, $\Lambda$ is a scalar field, and $\tau^{\mu}$ is a vector density. That is, the action Eq (\ref{htaction}) depends on the full unconstrained metric and the gauge symmetry includes the full diffeomorphism group of the manifold \cite{uqg2, ulqg2}. Variation with respect to $\Lambda$ yields the unimodular condition $\sqrt{-g} = \partial_{\mu}\tau^{\mu} $ while variation with respect to $\tau^{\mu}$ yields $\partial_{\mu}\Lambda = 0 $ which implies that $\Lambda$ is a constant in space and time - a cosmological constant. Now, variation with respect to the metric would yield the Einstein Field Equations (EFEs) with a non-zero cosmological constant $\Lambda$. This then suggests that any solution of the EFEs for any $\Lambda$ is a solution of the HT unimodular gravity theory.

We introduce the tetrad vectors $e^{\mu}_I$, where $e^{\mu}_Ie^{\nu}_Jg_{\mu\nu} = \eta_{IJ} = \mbox{diag}(-1,1,1,1)$, and the Lorentz Lie Algebra valued spin connection $\omega_{\mu}^{IJ}$ as independent variables (the so-called first order formulation \cite{book2}). The "spacetime" indices $(\mu,\nu,...)$ and the "internal" indices $(I,J,K,...)$ range as $\{0,1,2,3\}$. One can rewrite Eq (\ref{htaction}) in terms of the Ashtekar variables by adding a vanishing term (called the Holst term) and performing a $3+1$ split of spacetime $\mathcal{M} = \Sigma\times\mathbb{R}$ where $\Sigma(t)\equiv \Sigma_t$ is a compact spatial three-manifold evolving through an arbitrary choice of time $t\in \mathbb{R}$ which foliates $\mathcal{M}$ into the space-like Cauchy hypersurfaces. Upon partial gauge fixing to the time gauge, the inverse tetrads are $e_{\mu = a=\{1,2,3\}}^{I = 0} = 0$, $e_{\mu = 0}^{I = 0} = N$, $e_{\mu =0}^{I = i = \{1,2,3\}} = N^a e_a^i$, and $e_a^i$ (to be called the inverse triads of $\Sigma_t$), where $N$ and $N^a$ are the lapse and shift functions, respectively. The action $\mathcal{S}$ is then 
\begin{equation}\label{ashunim}
\mathcal{S} = \int dt \int_{\Sigma_t} d^3x \left[ \frac{1}{\kappa\gamma} E^a_i \partial_0 A_a^i +\frac{1 }{\kappa}\Lambda \partial_0 \tau^0 - \mathcal{H} \right]
\end{equation}
where,
\begin{equation}\label{fullh}
H = \int_{\Sigma } d^3x \mathcal{H} = \int_{\Sigma} d^3x \left[\Omega^iG_i + N^aV_a + \frac{1}{\kappa} \tau^a\partial_a \Lambda + N(S + \frac{1}{\kappa} \Lambda \sqrt{E} )\right]
\end{equation}
is the Hamiltonian, $A_a^i = \Gamma_a^i + \gamma \omega_a^{0i}$ is the su(2) valued Ashtekar-Barbero connection and $E^a_i = \mbox{det}(e^j_b)e^a_i$ is the densitized triad vector canonically conjugate to $A_a^i$. Moreover, $\Gamma_a^i = - \frac{1}{2}\epsilon^i_{jk}\omega_a^{jk}$ is the torsionless Levi-Civita connection on $\Sigma_t$ and $\omega_a^{0i}$ is the extrinsic curvature of $\Sigma_t$. The Lagrange multipliers $\Omega^i, N^a, \tau^a, N$ enforce the constraints $G_i = \partial_a E^a_i + \epsilon^k_{ij}A_a^jE^a_k = 0$ (called Gauss constraint as seen from gauge theories), $V_a = E^b_i \mathcal{F}^i_{ab} = 0$, $\partial_a \Lambda = 0$, and
\begin{equation}\label{scalcons}
\left(-\mathcal{F}^i_{ab} + (1+\gamma^2)\mathcal{R}^i_{ab}\right)\frac{\epsilon_i^{jk}E^a_jE^b_k}{2\kappa\gamma^2 \sqrt{E}} + \frac{\Lambda \sqrt{E}}{\kappa} = 0
\end{equation}
respectively, and where, $E = |\mbox{det}(E^a_i )|$, $\gamma$ is the real Barbero-Immirzi parameter introduced by the Holst term, $\mathcal{R}^i_{ab}$ is the curvature of $\Gamma_a^i$, and 
\begin{equation}\label{fullac}
\mathcal{F}^i_{ab}  = \partial_a A_b^i - \partial_b A_a^i + \epsilon^i_{jk}A_a^jA_b^k
\end{equation}
is the curvature of $A_a^i$. Since the Hamiltonian is just a linear sum of the vanishing constraints, we then have a vanishing Hamiltonian $H = 0$.

Note that since $\partial_a \Lambda = 0$, we see that $T = \int_{\Sigma_t} d^3x \tau^0$ is canonically conjugate to $\Lambda$. It can be interpreted by considering the integration of the unimodular condition over a region $\mathcal{C} \subset \mathcal{M}$ bounded by $\Sigma_{t_2}$ and $\Sigma_{t_1}$
\begin{equation}
\int_{\mathcal{C}}\sqrt{-g} = \int_{\mathcal{C}} \partial_{\mu} \tau^{\mu} = \int_{t_1}^{t_2} dt \int_{\Sigma_t} d^3x \partial_{0} \tau^{0} = T(t_2) - T(t_1)
\end{equation}
which is just the four-volume between the hypersurfaces $\Sigma_{t_2}$ and $\Sigma_{t_1}$. $T$ can be considered as a "cosmological time" since it increases continuously for any future directed time-like curve \cite{ulqc}. Also note that $T$ is still off-shell since we have not used all the constraints yet.

\section{Reduction to a Cosmological Model}\label{cosm}
We now consider a finite, spatially flat, homogeneous and isotropic patch $\mathcal{C}$ of spacetime described by the metric 
\begin{equation}
ds^2 = g_{\mu\nu}dx^{\mu}dx^{\nu} = -dt^2 + a^2(t) \; ^{\circ}q_{ab} dx^adx^b
\end{equation}
where, $^{\circ}q_{ab}$ is a constant fiducial metric on $\Sigma_t$ which we consider to be topologically equivalent to $\mathbb{R}^3$, and $a(t)$ is the scale factor characterizing the size of the finite space-like cell $\Sigma_t$ of $\mathcal{C} = \Sigma\times\mathbb{R}$ at some proper time $t$. Namely, $a^3(t)V_0$ is the physical (three-dimensional) volume of $\Sigma_t$ at $t$, where $V_0 = \int_{\Sigma}d^3x$ is the nondynamical coordinate volume. Because the space is flat, homogeneous, and isotropic, we can let the cell $\Sigma_t$ be cubical with respect to every physical metric on $\mathbb{R}^3$ \cite{lqcrev3}. We consider the case where there is a nonzero cosmological constant $\Lambda$ and for simplicity, a vanishing stress-energy tensor so that the scale factor satisfies the Friedmann Equations (the components of the EFEs)
\begin{equation}\label{fe}
\frac{\dot{a}^2}{a^2} = \frac{\Lambda}{3} \;\;\;\;\;\;\;\;,\;\;\;\;\;\;\;\;  \frac{\ddot{a}}{a}= \frac{\Lambda}{3}
\end{equation}
Note that they imply $a_{\pm}(t) = a_0 \exp\left( \pm \sqrt{\frac{\Lambda}{3}} t\right)$ where  $a_+(t)$ ($a_-(t)$) represents an always expanding (contracting) $\Sigma_t$, In symbols, $\dot{a}_+>0$ ($\dot{a}_-<0$) for all $t$. 
This system can be equivalently described in the Hamiltonian formulation by considering the Ashtekar variables
\begin{equation}\label{acep}
A_a^i = c(t) V_0^{-1/3} \; ^{\circ}\omega_a^i  \;\;\;\;\;\;\;\;,\;\;\;\;\;\;\;\; E^a_i = p(t) V_0^{-2/3} \sqrt{\mbox{det}(^{\circ}q_{ab})} \; ^{\circ}e^a_i 
\end{equation}
where $ ^{\circ}e^a_i $ is a fiducial triad ($^{\circ}e^a_i \; ^{\circ}e^b_j \; ^{\circ}q_{ab} = \delta_{ij}$) which can be diagonalized and $^{\circ}\omega_a^i$ is its inverse. The only dynamical factors are $c(t) =  \gamma \dot{a} V_0^{1/3}$ and $p(t) = \pm a^2 V_0^{2/3}$. Note that the connection $c(t)$ represents only the extrinsic curvature of $\Sigma_t$ since the Levi-Civita connection $\Gamma_a^i$ vanishes on the flat hypersurface. Moreover, $p(t)$ represents the magnitude of the triad vectors so that $\mbox{sgn}(p)$ represents their orientation with respect to the fiducial triad vectors $ ^{\circ}e^a_i $. Classically, this does not change for the underlying manifold to be orientable. In the quantum theory however, this can be allowed to change. The conjugate pair $(c,p)$ satisfies the Poisson Bracket
\begin{equation}\label{pbcp}
\left\{ c,p\right\} = \frac{\kappa\gamma}{3}
\end{equation}
which can be read off from the action, Eq (\ref{ashunim}). It can be shown that the Ashtekar variables given by Eq (\ref{acep}) already satisfy the constraints $G_i = 0$ and $V_a = 0$ (due to homogeneity and isotropy as well). Along with the spatial constancy of $\Lambda$, the only vanishing constraint left to impose is Eq (\ref{scalcons}) which can be further simplified. Using Eq (\ref{acep}), the curvature of the connection $A_a^i \sim c(t)$, given fully by Eq (\ref{fullac}), reduces to 
\begin{equation}\label{curvc}
\mathcal{F}^i_{ab} = c^2(t) V_0^{-2/3} \epsilon^i_{jk} \;^{\circ}\omega_a^j \; ^{\circ}\omega_b^k
\end{equation}
and owing to the spatial flatness of $\Sigma_t$, the curvature $\mathcal{R}^i_{ab}$ of $\Gamma_a^i$ vanishes so that the Hamiltonian (Eq (\ref{fullh}) with $N=1$ so that $t$ is the proper time) reduces to\begin{equation}\label{hcons}
H = -\frac{3c^2 |p|^{1/2}}{\kappa\gamma^2} + \frac{\Lambda |p|^{3/2}}{\kappa}
\end{equation}
And so, we can impose the Hamiltonian constraint as the vanishing of Eq (\ref{hcons}). Using $H=0$, along with the equations of motion $\dot{p} = \{p,H\}$ and $\dot{c} = \{c,H\}$, the Friedmann Equations, Eqs (\ref{fe}), are reproduced.  Additionally, $(T, \Lambda)$ form another pair of dynamical conjugated variables satisfying the Poisson Bracket
\begin{equation}\label{pbtl}
\left\{ T, \Lambda \right\} = \kappa
\end{equation}
which also can be read off from Eq (\ref{ashunim}). One readily finds that $\dot{\Lambda} = \{\Lambda,H\} = 0$ so that $\Lambda$ is also a constant in time and $\dot{T}(t) = \{T,H\} = |p(t)|^{3/2} = a^3(t)V_0$ is the three-volume of $\Sigma_t$. Integrating, we see that $T(t) = \int_0^{t}|p(t')|^{3/2}dt'$ is indeed a four-volume and can be interpreted as the four-volume preceding the hypersurface $\Sigma_t$. Also note that $T(t)$ is explicitly monotonic for any timelike curve so that, indeed, it can be called as a cosmological time.

We wish to find an expression for $T$ using only variables on $\Sigma_t$. We proceed first by imposing the vanishing of the Hamiltonian Eq (\ref{hcons}), to arrive at an expression for $\Lambda$ on-shell which we denote as $\tilde{\Lambda}$. Namely,
\begin{equation}\label{classL}
\tilde{\Lambda} = \frac{3c^2}{\gamma^2 |p|}
\end{equation}
and use the equations of motion $\dot{\Lambda}  = 0$, $\dot{T}= |p|^{3/2} $, and $\gamma \dot{p} = 2c|p|^{1/2}$, so that
\begin{equation}
\tilde{\Lambda}(c,p) = \tilde{\Lambda}(c',p') = \frac{3c'^2}{\gamma^2 |p'|} = \frac{3\dot{p'}^2}{4|p'|^2} = \frac{3|p'|\dot{p'}^2}{4\dot{T'}^2}
\end{equation}
where the unprimed variables are evaluated at the hypersurface $\Sigma_t$ while the primed variables are evaluated at some other hypersurface $\Sigma_{t'}$ where $t' \neq t$. Taking the square root, we get
\begin{equation}
\frac{\sqrt{3}}{2}|p'|^{1/2}\frac{dp'}{dT'} = \pm \sqrt{\tilde{\Lambda}(c,p)} = \pm\frac{\sqrt{3}|c|}{\gamma |p|^{1/2}} = \frac{\sqrt{3}c}{\gamma |p|^{1/2}}
\end{equation}
while remembering that the positive (negative) sign, and so $c=\gamma \dot{a}>0$ ($c=\gamma \dot{a}<0$), represents the expanding (contracting) hypersurface $\Sigma_t$. Integrating,
\begin{eqnarray}\label{classT}
\int_0^{\tilde{T}}dT' = \left(\pm \frac{\sqrt{3}}{2\sqrt{\tilde{\Lambda}(c,p)}} \right)\int_p^0 |p'|^{1/2} dp' = \left(\frac{\gamma |p|^{1/2}}{2c}\right) \frac{2}{3}\mbox{sgn}(p)\left(0-|p|^{3/2}\right) \nonumber\\
\tilde{T}(c,p) = - \mbox{sgn}(p)\frac{\gamma |p|^2}{3c}
\end{eqnarray}
so that $\tilde{T}(c,p)$ is the four-volume preceding $\Sigma_{t'}$ with volume $|p(t')|^{3/2} = 0$ and succeeding $\Sigma_{t}$ with volume $|p(t)|^{3/2}$. As with $\tilde{\Lambda}$, we denote Eq (\ref{classT}) with a tilde since it is now a quantity evaluated on-shell. Note that for $p>0$, $\tilde{T} < 0$ ($\tilde{T} > 0$) when $c>0$ ($c<0$) so that $\Sigma_{t'}$ happened before (after) $\Sigma_{t}$ for an expanding (contracting) hypersurface. One can also calculate that Eq (\ref{classT}) still satisfies the Poisson Bracket Eq (\ref{pbtl}). Thus, we interpret $\tilde{T}(c,p)$ as indeed measuring the volume of a finite four-dimensional region of spacetime.

\section{The Hilbert Space Properties of Loop Quantum Cosmology}\label{hilb}
Having the conjugate pairs $(c,p)$ and $(T,\Lambda)$, constructing the quantum theory should be straightforward by mapping the pairs into operators acting on the kinematical Hilbert space $\mathcal{H}^{kin}$ of Loop Quantum Cosmology (LQC) and their Poisson Brackets to commutation relations. However, according to Loop Quantum Gravity (LQG), the connection $c$ has no corresponding operator on $\mathcal{H}^{kin}$ \cite{lqcart1} so that the quantities with geometrical operator analogs are the holonomies of the connection. That is, the path ordered exponential of the connection along a curve. Due to the homogeneity and isotropy of the space we are considering, it suffices to consider holonomies along straight line segments with coordinate length $\mu V_0^{1/3}$ parallel to the $j^{th}$ edge of the cell $\Sigma_t$, where $V_0^{1/3}$ is the coordinate length of the $j^{th}$ edge. The holonomy of the connection along such an edge is \cite{lqcrev1}
\begin{equation}
h_j^{(\mu)} = \exp(\mu c \tau_j) = \cos\left(\frac{\mu c}{2} \right) \mathbb{I} + 2 \sin\left(\frac{\mu c}{2} \right)\tau_j
\end{equation}
which can be written in terms of the complex exponentials $\tilde{\mathcal{N}}_{\mu} = \exp\left( i\mu c/2\right)$. Also note that $\mathbb{I}$ is the $2\times 2$ identity matrix, and $2i\tau_j = \sigma_j$ are the spin Pauli matrices. It is then the holonomies $h_j^{(\mu)}$ (or equivalently, $\tilde{\mathcal{N}}_{\mu} $) which can be operators acting on $\mathcal{H}^{kin}$. Now, the physical Hilbert space $\mathcal{H}^{phys}$ of LQC is the space of states which are annihilated by the operator version of the Hamiltonian constraint Eq (\ref{hcons}) \cite{lqcrev1}. That is, for $\Phi_{phys} \in \mathcal{H}^{kin}$ satisfying $\hat{H}\Phi_{phys} = 0$ then $\Phi_{phys} \in \mathcal{H}^{phys}$. In order to construct an operator version $\hat{H}$ of Eq (\ref{hcons}) however, one needs to rewrite the curvature $\mathcal{F}^i_{ab}$, given by Eq (\ref{curvc}), in terms of $\tilde{\mathcal{N}}_{\mu}$. This can be done by considering holonomies $h_{\square_{ij}}^{(\mu)}$ around a square loop $\square_{ij}$ in the $i-j$ plane spanned by a face of the cell $\Sigma_t$ where each side of $\square_{ij}$ has coordinate length $\mu V_0^{1/3}$ so that $h_{\square_{ij}}^{(\mu)} = h_i^{(\mu)}h_j^{(\mu)}h_i^{(-\mu)}h_j^{(-\mu)}$ \cite{lqcrev1}. The curvature $\mathcal{F}^i_{ab}$ can then be written in terms of $h_{\square_{ij}}^{(\mu)}$ as the square loop $\square_{ij}$ (or, equivalently $\mu$) approaches zero. However, we cannot let the area of $\square_{ij}$ go to zero in the quantum theory \cite{lqcrev2}. Instead, we let the area of the loop shrink until it has a minimum coordinate area of $\bar{\mu}^2V_0^{2/3}$ such that the physical area corresponding to this is $a^2\bar{\mu}^2V_0^{2/3} = |p|\bar{\mu}^2$ which we set to be equal to $\Delta$, the minimum eigenvalue of the area operator in full LQG \cite{art1}. This prescription, $|p|\bar{\mu}^2 = \Delta$ is called improved dynamics \cite{lqcbb3}. Then, Eq (\ref{curvc}) becomes \cite{lqcrev1}
\begin{equation}
\mathcal{F}^i_{ab} = -2 \lim_{\mu\rightarrow\bar{\mu}}\mbox{Tr}\left(\frac{(h_{\square_{ij}}^{(\mu)} - \mathbb{I})}{\mu^2 V_0^{2/3}}\tau^i\right) \;^{\circ}\omega_a^j \; ^{\circ}\omega_b^k = \frac{\sin^2(\bar{\mu}c)}{\bar{\mu}^2}\frac{\epsilon^i_{jk} \;^{\circ}\omega_a^j \; ^{\circ}\omega_b^k}{V_0^{2/3} }
\end{equation}
so that we effectively replace 
\begin{equation}\label{crep}
c^2 \mapsto \frac{\sin^2(\bar{\mu}c)}{\bar{\mu}^2} = - \frac{|p|}{4\Delta}\left(\tilde{\mathcal{N}}_{2\bar{\mu}} - \tilde{\mathcal{N}}_{-2\bar{\mu}}  \right)^2
\end{equation}
which now have operator analogs. This suggests that the curvature is non-local \cite{lqcrev3}. Note that we recover the original curvature $c^2$ if we insist on $\bar{\mu}\rightarrow 0$. We introduce a change in variables \cite{lqcrev1},
\begin{equation}\label{chvar}
b = \frac{\hbar \bar{\mu} c}{2}  =\frac{\hbar \sqrt{\Delta}}{2|p|^{1/2}} c \;\;\;\;\;\;\;\;,\;\;\;\;\;\;\;\; \nu = \frac{\mbox{sgn}(p)|p|^{3/2}}{2\pi\gamma \ell^2_{pl} \sqrt{\Delta}}
\end{equation}
with $\ell_{pl} = \sqrt{G\hbar}$ as the Planck length so that $\tilde{\mathcal{N}}_{\lambda\bar{\mu}}  = \exp\left(i \lambda b/ \hbar \right) \equiv \mathcal{N}_{\lambda}$. Then, from Eq (\ref{pbcp}), we get the Poisson Bracket $\{b,\nu\} = 1$ which, in turn, gives $\{\mathcal{N}_{\lambda} ,\nu\} = \frac{i\lambda}{\hbar}\mathcal{N}_{\lambda}$. Additionally, one can see that $|\nu|$ is proportional to the volume. Thus, the basic operators acting on $\mathcal{H}^{kin}$ are $\hat{\mathcal{N}}_{\lambda}$ and $\hat{\nu}$ and their Poisson Bracket structure is mapped into an operator commutation relation
\begin{equation}\label{hvccr}
\left[ \hat{\mathcal{N}}_{\lambda}, \hat{\nu} \right] = i\hbar\widehat{ \{\mathcal{N}_{\lambda} ,\nu\} } = - \lambda \hat{\mathcal{N}}_{\lambda}
\end{equation}

The kinematic Hilbert space $\mathcal{H}^{kin}$ is then $\mathcal{L}^2(\overline{\mathbb{R}}_{Bohr},d\mu_{Bohr})$, the space of square integrable functions on the Bohr compactification of the real line $\overline{\mathbb{R}}_{Bohr}$ equipped with the Haar measure $\mu_{Bohr}$ \cite{lqcart1}. That is, it is the space of normalizable states $\Phi(b)\in \mathcal{H}^{kin}$ under the inner product
\begin{equation}\label{innpr}
\left<\Phi_1 | \Phi_2 \right> = \lim_{D\rightarrow \infty} \frac{1}{2D}\int_{-D}^D \Phi_1^*(b)\Phi_2(b) db
\end{equation}
with a basis given by
\begin{equation}
\mathcal{N}_{\nu}(b) = \exp\left(i\frac{\nu b}{\hbar}\right)
\end{equation}
The set of basis states $\mathcal{N}_{\nu}(b)$ form a complete and non-countable orthonormal set
\begin{equation}
\left<\mathcal{N}_{\nu_1} | \mathcal{N}_{\nu_2}\right> = \delta_{\nu_1\nu_2}
\end{equation}
so that the Hilbert space $\mathcal{H}^{kin}$ is non-separable. In fact, the $\mathcal{N}_{\nu}(b)$'s are the LQC analogs of the spin network functions in full LQG \cite{art6}. A general state $\Phi(b)\in\mathcal{H}^{kin}$ is then a countable linear combination
\begin{equation}
\Phi(b) = \sum_{\nu}\Psi(\nu)\mathcal{N}_{\nu}(b) = \sum_{\nu}\Psi(\nu) \exp\left(i\frac{\nu b}{\hbar}\right)
\end{equation}
with $||\Phi ||^2 =\left<\Phi | \Phi \right> = \sum_{\nu} |\Psi(\nu)|^2<\infty$. They are sometimes called almost periodic functions because the $\nu$'s are allowed to be arbitrary real numbers rather than integer multiples of a fixed number. Note that the normalizable state vectors are not integrals but a discrete sum. Consequently, the intersection between $\mathcal{H}^{kin}$ and the usual Hilbert space $\mathcal{L}^2(\mathbb{R},db)$ of quantum mechanics is the zero vector \cite{lqcrev3}. Thus, the structure of the LQC Hilbert space is very different from the Hilbert space used in usual Quantum Cosmology - the Wheeler-DeWitt (WDW) theory. 

The actions of $\hat{\mathcal{N}}_{\lambda}$ and $\hat{\nu}$ on an arbitrary state $\Phi(b)\in \mathcal{H}^{kin}$ which respects the commutator structure of Eq (\ref{hvccr}) are given by 
\begin{equation}
\hat{\mathcal{N}}_{\lambda}\Phi(b) = \mbox{e}^{i\lambda b /\hbar }\Phi(b) \;\;\;\;\;\;\;\;,\;\;\;\;\;\;\;\; \hat{\nu}\Phi(b) = \frac{\hbar}{i}\frac{d}{db}\Phi(b)
\end{equation}
respectively. Note that the following operations hold: $\hat{\mathcal{N}}_{\lambda}\hat{\mathcal{N}}_{\lambda'} = \hat{\mathcal{N}}_{\lambda'}\hat{\mathcal{N}}_{\lambda}  = \hat{\mathcal{N}}_{\lambda + \lambda'}$, $\hat{\mathcal{N}}_{0} = \hat{\mathbb{I}}$, and $(\hat{\mathcal{N}}_{\lambda})^{-1} = \hat{\mathcal{N}}_{-\lambda}$, where $\hat{\mathbb{I}}$ is the identity operator on $\mathcal{H}^{kin}$. One can also readily calculate their actions on the basis as 
\begin{equation}
\hat{\mathcal{N}}_{\lambda}\mathcal{N}_{\nu}(b) = \mathcal{N}_{\nu + \lambda}(b)  \;\;\;\;\;\;\;\;,\;\;\;\;\;\;\;\; \hat{\nu}\mathcal{N}_{\nu}(b) = \nu\mathcal{N}_{\nu}(b)  
\end{equation}
so that the $\mathcal{N}_{\nu}(b)$'s are the normalizable eigenstates of $\hat{\nu}$ with eigenvalue $\nu$. Their normalizability suggests that the spectrum of $\hat{\nu}$ (which is proportional to the volume operator) is discrete. This is what one means when one says that the (spatial three-dimensional) volume is discrete \cite{lqcart1}. Moreover, one can say that the $\mathcal{N}_{\nu}(b)$'s are states of definite volume $\nu$. 
We now pose a similar question for the four-volume. That is, if one can construct an operator acting on $\mathcal{H}^{kin}$ which corresponds to the spacetime four-volume $\tilde{T}(c,p)$, Eq (\ref{classT}), is its spectrum discrete?

\section{A Quantum Spacetime Four-Volume Operator}\label{Top}
We are now considering the quantum version of the classical system in Section \ref{cosm}. That is, our quantum system is now the finite space-like hypersurface $\Sigma_t$ evolving through the proper time $t$ without matter sources and with non-vanishing $\Lambda$. We wish to find an operator version of Eq (\ref{classT}) acting on $\mathcal{H}^{kin}$. However, $\tilde{T}(c,p)$ contains a factor of $c^{-1}$ which as stated earlier has no operator on $\mathcal{H}^{kin}$ corresponding to it. One needs to replace it in terms of the holonomies much like Eq (\ref{crep}) which can lead us to a replacement like $c^{-1} \mapsto \bar{\mu}(\sin(\bar{\mu}c))^{-1}$. How can one determine if this is correct, however? We impose that such an operator must still satisfy the canonical commutation relation
\begin{equation}\label{ccrtl}
\left[ \hat{T}, \hat{\Lambda} \right] = i\hbar \widehat{\left\{ \tilde{T},\tilde{\Lambda}  \right\}}  = i\hbar\kappa \hat{\mathbb{I}} = 8\pi i \ell^2_{pl} \hat{\mathbb{I}}
\end{equation}
which just comes from mapping the Poisson Bracket Eq (\ref{pbtl}) into a commutator. However, the operator constructed from such a replacement does not satisfy Eq (\ref{ccrtl}). Instead, we consider the commutator Eq (\ref{ccrtl}) as the starting point to determine a four-volume operator $\hat{T}$. This idea of solving the commutation relation when quantization fails to map the classical Poisson Bracket algebra into a quantum operator commutator algebra originated from the idea of \emph{supraquantization} \cite{supq}. However, we first need to construct an operator $ \hat{\Lambda}$ for $\tilde{\Lambda}(c,p)$ given by Eq (\ref{classL}). The factor of $c^2$ can still be interpreted as the curvature, Eq (\ref{curvc}), of the connection $c$ so that we still use the replacement Eq (\ref{crep}) 
\begin{equation}
\tilde{\Lambda} = \frac{3}{\gamma^2 |p|}c^2 \mapsto \tilde{\Lambda} = \frac{3}{\gamma^2 |p|}\left( - \frac{|p|}{4\Delta}\left(\mathcal{N}_{2} - \mathcal{N}_{-2}  \right)^2
 \right)
\end{equation}
which now have operator analogs. We consider the case where the factors of $|p| \propto |\nu|^{2/3} $ do not contribute to the quantum theory (since they cancel classically). That is, we suppose that there are no inverse volume corrections in the quantum theory for now. Their contributions can be studied in future works. Going back, we get the operator
\begin{equation}\label{qL}
\hat{\Lambda} = -\frac{3}{4\gamma^2 \Delta}\left(\hat{\mathcal{N}}_{2} - \hat{\mathcal{N}}_{-2}  \right)^2 = -\frac{3}{4\gamma^2 \Delta} \left(\hat{\mathcal{N}}_{4} + \hat{\mathcal{N}}_{-4} -2 \hat{\mathbb{I}}\right)
\end{equation}
So, given $\hat{\Lambda}$ and Eq (\ref{ccrtl}), we intend to construct $\hat{T}$. However, this would not yield a unique operator since we can add any function $f(\hat{\Lambda})$ to $\hat{T}$ so that $\hat{T} + f(\hat{\Lambda})$ still satisfies Eq (\ref{ccrtl}). An interesting direction was suggested where perhaps one could find a function $f(\hat{\Lambda})$ such that $\hat{T} + f(\hat{\Lambda})$ is self-adjoint in $\mathcal{H}^{kin}$. This may not be straightforward however as, for a given domain, there may be unique, multiple, or even no function $f(\hat{\Lambda})$ where $\hat{T} + f(\hat{\Lambda})$ is self-adjoint. Or perhaps one can go in the other direction where one makes an ansatz for $f(\hat{\Lambda})$ and construct a domain where $\hat{T} + f(\hat{\Lambda})$ is self-adjoint. Once again, there may be a unique, multiple, or an empty domain for such a condition. This interesting direction can be tackled in future studies. Nevertheless, we proceed by assuming an expansion for $\hat{T}$ 
\begin{equation}\label{asT}
\hat{T} = \sum_{m,n}\alpha_{m,n} \hat{T}_{m,n} 
\end{equation}
where the $\alpha_{m,n}$'s are to be determined and the $\hat{T}_{m,n}$'s are similar to the Bender-Dunne basis operators \cite{art7, art8} whose status as Hilbert Space operators were studied in \cite{bu1}. Specifically, we let
\begin{equation}\label{tmn}
\hat{T}_{m,n} = \frac{1}{2^n}\sum_{k=0}^{\infty}\frac{n!}{k!(n-k)!}\hat{\nu}^k\hat{\mathcal{N}}_{m} \hat{\nu}^{n-k} = \frac{1}{2^n}\sum_{k=0}^{\infty}\frac{n!}{k!(n-k)!}\hat{\nu}^{n-k}\hat{\mathcal{N}}_{m} \hat{\nu}^{k} 
\end{equation}
where, $m,n,k$ are positive integers. Note that $\hat{T}_{0,0} = \hat{\mathbb{I}}$. However, since $\lambda$ in $\hat{\mathcal{N}}_{\lambda} $ can be any real number, perhaps a possible extension of $\hat{T}_{m,n}$ can be made in a future study. Referring to \ref{consT}, an operator $\hat{T}$ constructed from such a process is
\begin{equation}\label{quanT}
\hat{T} = - \frac{i\hbar \kappa\gamma^2 \Delta}{6} \left[ \left( \hat{\mathcal{N}}_4 - \hat{\mathcal{N}}_{-4} \right)^{-1}\hat{\nu}  + \hat{\nu} \left( \hat{\mathcal{N}}_4 - \hat{\mathcal{N}}_{-4} \right)^{-1} \right]
\end{equation}
A similar process was used in \cite{bu2, bu3} to construct Relativistic Time of Arrival Operators from their canonical commutation relations with the system Hamiltonians. 

We go to the classical limit by "removing the hats" (letting $\hat{T}$ act on a definite volume state $\mathcal{N}_{\nu}(b)$) and let $\Delta \rightarrow 0$. Specifically,
\begin{eqnarray*}
\hat{T} \rightarrow - \frac{i\hbar \kappa\gamma^2 \Delta}{3}\frac{\nu}{\mathcal{N}_4 - \mathcal{N}_{-4}} &= - \frac{\hbar \kappa\gamma^2 \Delta}{6} \nu \frac{2i}{\mbox{e}^{i4b/\hbar} - \mbox{e}^{-i4b/\hbar} } \\
&= - \frac{\gamma\sqrt{\Delta}}{3}2\mbox{sgn}(p)|p|^{3/2}\frac{2i}{\mbox{e}^{i2\bar{\mu}c} - \mbox{e}^{-i2\bar{\mu}c} }\\
&= - \frac{\gamma}{3} \mbox{sgn}(p)|p|^2\frac{2\sqrt{\Delta}|p|^{-1/2} }{\sin\left( 2 \sqrt{\Delta}|p|^{-1/2} c\right)}\\
& \rightarrow - \frac{\gamma \; \mbox{sgn}(p)|p|^2}{3 c} = \tilde{T}(c,p)
\end{eqnarray*}
We can also construct $\hat{T}$ from $\tilde{T}(c,p)$ by replacing the connection with the holonomies as $c^{-1}\rightarrow 2\bar{\mu}(\sin(2\bar{\mu}c))^{-1} $ and then mapping the holonomies into operators. Note that $c^{-1}$ is recovered if we insist on $\bar{\mu}\rightarrow 0$ ($\Delta \rightarrow 0$). Although the motivation for such a replacement is unknown to the author. Nevertheless, the recovery of $\tilde{T}(c,p)$ from $\hat{T}$ and its conjugacy with $\hat{\Lambda}$ (by construction) suggests that $\hat{T}$ is indeed an operator corresponding to the spacetime four-volume between a space-like three-dimensional hypersurface with non-vanishing volume and the hypersurface with vanishing volume it evolves into or evolves from. Put simply then, $\hat{T}$ can indeed be interpreted as a quantum spacetime four-volume operator.

\section{ The Discrete Spectrum of $\hat{T}$ }\label{disc}
In this section, we will calculate the point or discrete spectrum of $\hat{T}$ which we will denote by $\sigma_p(\hat{T})$. By definition, $\sigma_p(\hat{T})$ is the set of all $\tau$'s in which the resolvent operator $(\tau \hat{\mathbb{I}} - \hat{T})$ is not invertible. It turns out that $\sigma_p(\hat{T}) = \mathbb{C}$ so that the spectrum of $\hat{T}$ is purely discrete and is the entire complex plane. That is, we will show that the equation 
\begin{equation}\label{noninv}
(\tau  \hat{\mathbb{I}} - \hat{T}) \Phi_{\tau} = 0
\end{equation}
has a solution $\Phi_{\tau}$ in the domain of $\hat{T}$ for all $\tau \in \mathbb{C}$ so that $(\tau \hat{\mathbb{I}} - \hat{T})$ is not invertible for any complex number $\tau$. We may assign the maximal domain 
\begin{equation}\label{maxdom}
D(\hat{T}) = \{ \Phi \in \mathcal{H}^{kin} \left| ||\hat{T}\Phi ||^2 < \infty \right.\}
\end{equation}
for our operator $\hat{T}$.

To find the formal solutions for Eq (\ref{noninv}), we first calculate the action of $\hat{T}$ on an arbitrary $\Phi(b) \in \mathcal{H}^{kin}$. From Eq (\ref{quanT}), we have
\begin{eqnarray}
\hat{T}\Phi(b) &= - \frac{i\hbar \kappa\gamma^2 \Delta}{6} \left[ \left( \mbox{e}^{i4b/\hbar} - \mbox{e}^{-i4b/\hbar} \right)^{-1}\frac{\hbar}{i}\frac{d\Phi}{db}  + \frac{\hbar}{i}\frac{d}{db}  \left( \frac{\Phi}{\mbox{e}^{i4b/\hbar} - \mbox{e}^{-i4b/\hbar} } \right) \right] \nonumber\\
&= \frac{i\hbar^2 \kappa\gamma^2 \Delta}{12} \left[ \frac{1}{\sin(4b/\hbar)}\frac{d\Phi}{db} +  \frac{d}{db}  \left( \frac{\Phi}{\sin(4b/\hbar)} \right) \right] \nonumber\\
&= \frac{i\hbar^2 \kappa\gamma^2 \Delta}{12} \left[ \frac{2}{\sin(4b/\hbar)}\frac{d\Phi}{db} -  \frac{1}{\sin^2(4b/\hbar)}  \cos\left(\frac{4b}{\hbar}\right) \frac{4}{\hbar}\Phi \right] \nonumber\\
&= \frac{i\hbar \kappa\gamma^2 \Delta}{6} \left[ \frac{\hbar}{\sin(4b/\hbar)}\frac{d\Phi}{db} -  2 \frac{\cot(4b/\hbar)}{\sin(4b/\hbar)}  \Phi \right]
\end{eqnarray}
so that Eq (\ref{noninv}) becomes
\begin{eqnarray}
\;\;\;\;\;\;\;\;\;\;\;\;\;\;\;\; \;\;\;\;\;\;\;\;\;\;\;\;\;\;\;\;\;\;\;\;\;\;\;\;\;\;\; \hat{T}\Phi_{\tau}(b) &= \tau  \Phi_{\tau}(b)  \nonumber\\
\;\;\;\frac{i\hbar \kappa\gamma^2 \Delta}{6} \left[ \hbar \frac{d\Phi_{\tau}}{db} - 2 \cot\left(\frac{4b}{\hbar}\right)\Phi_{\tau} \right] &= \tau\sin\left(\frac{4b}{\hbar}\right)\Phi_{\tau} \nonumber\\
\;\;\;\;\;\;\;\;\;\;\;\;\;\;\;\;\;\;\; \hbar \frac{d\Phi_{\tau}}{db} - 2 \cot\left(\frac{4b}{\hbar}\right)\Phi_{\tau}  &= \frac{2\rho_{crit}\tau}{i\hbar} \sin \left(\frac{4b}{\hbar}\right)\Phi_{\tau}\nonumber\\
\ln\left(\left|\frac{\Phi_{\tau}(b)}{\Phi_0}\right|\right) - \frac{1}{2}\ln\left(\left| \sin\left(\frac{4b}{\hbar}\right)\right|\right) &= i \frac{\rho_{crit}\tau}{2\hbar}\cos \left(\frac{4b}{\hbar}\right) 
\end{eqnarray}
which gives us the formal solutions
\begin{equation}\label{eigT}
\Phi_{\tau}(b) = \Phi_0 \left|\sin\left(\frac{4b}{\hbar}\right)\right|^{1/2} \exp\left( \frac{i\rho_{crit}\tau}{2\hbar} \cos \left(\frac{4b}{\hbar}\right) \right) 
\end{equation}
for each $\tau \in \mathbb{C}$. We let $\Phi_0$ to be some constant and 
\begin{equation}
\rho_{crit} = \frac{3}{\kappa\gamma^2\Delta}
\end{equation}
to be the maximum energy density of matter sources from LQC \cite{lqcrev3}. For our study, perhaps this only serves as a placeholder as we do not have any matter source for our system.

Having calculated the formal solutions $\Phi_{\tau}(b)$ for Eq (\ref{noninv}), we now show that for all complex number $\tau$'s, all of the corresponding $\Phi_{\tau}(b)$'s are indeed in $D(\hat{T})$ given by Eq (\ref{maxdom}). That is, we should have $||\hat{T}\Phi_{\tau}||^2 = |\tau|^2  \; ||\Phi_{\tau}||^2 < \infty $. In other words, $\Phi_{\tau}$ must be normalizable in $\mathcal{H}^{kin}$ for all $\tau \in \mathbb{C}$. We first calculate the norm of the $\Phi_{\tau}(b)$'s given by Eq (\ref{eigT}) for $\tau\in\mathbb{R}$.
\begin{eqnarray}
|| \Phi_{\tau\in\mathbb{R}} ||^2 &= \left< \Phi_{\tau\in\mathbb{R}} | \Phi_{\tau\in\mathbb{R}}\right> = \lim_{D' \rightarrow\infty} \frac{1}{2D'}\int_{-D'}^{D'} \Phi^*_{\tau\in\mathbb{R}} (b) \Phi_{\tau\in\mathbb{R}} (b) db \nonumber\\
&= \lim_{D'\rightarrow\infty} \frac{1}{2D'}\int_{-D'}^{D'}  |\Phi_0 ^{(\mathbb{R})}|^2 \left|\sin\left(\frac{4b}{\hbar}\right)\right| db \nonumber\\
&=  |\Phi_0 ^{(\mathbb{R})}|^2 \lim_{D'\rightarrow\infty} \frac{1}{D'}\int_{0}^{D'}  \left|\sin\left(\frac{4b}{\hbar}\right)\right| db  \nonumber\\
&=  |\Phi_0 ^{(\mathbb{R})} |^2 \lim_{D\rightarrow\infty} \frac{1}{D}\int_{0}^{D}  \left|\sin\left(x\right)\right| dx  \nonumber\\
&=  |\Phi_0 ^{(\mathbb{R})} |^2 \lim_{ \mathbb{N}\ni n \rightarrow\infty} \frac{1}{n\pi} \sum_{j=0}^{n-1}\int_{j\pi}^{(j+1)\pi} \left|\sin\left(x\right)\right| dx\nonumber\\
&=  |\Phi_0 ^{(\mathbb{R})} |^2 \lim_{ \mathbb{N}\ni n \rightarrow\infty} \frac{1}{n\pi} \sum_{j=0}^{n-1}\int_{0}^{\pi} \left|\sin\left(x+j\pi\right)\right| dx\nonumber\\
&=  |\Phi_0 ^{(\mathbb{R})} |^2 \lim_{ \mathbb{N}\ni n \rightarrow\infty} \frac{1}{n\pi} \sum_{j=0}^{n-1}\int_{0}^{\pi} \left|(-1)^j \sin\left(x\right)\right| dx\nonumber\\
&=  |\Phi_0 ^{(\mathbb{R})} |^2 \lim_{ \mathbb{N}\ni n \rightarrow\infty} \frac{1}{n\pi} \sum_{j=0}^{n-1}\int_{0}^{\pi} \sin\left(x\right)dx\nonumber\\
&=  |\Phi_0 ^{(\mathbb{R})} |^2 \lim_{ \mathbb{N}\ni n \rightarrow\infty} \frac{1}{n\pi}  2n =  |\Phi_0 ^{(\mathbb{R})} |^2 \frac{2}{\pi} <\infty\nonumber\\
\end{eqnarray}
Where, we let $x = 4b/\hbar$ and $D = 4D'/\hbar$ in the fourth line. Thus, the $\Phi_{\tau\in\mathbb{R}}$'s are normalizable in $\mathcal{H}^{kin}$ and in turn, shows that all $\Phi_{\tau\in\mathbb{R}} \in D(\hat{T}) $ are eigenstates of $\hat{T}$. Moreover, they are normalized if we set $\Phi_0 ^{(\mathbb{R})} = \sqrt{\pi/2}$. The real line is then in the discrete spectrum of $\hat{T}$. Similarly, we can proceed to the $\Im[\tau]\neq 0$ case, where $\Im[\tau]$ is the imaginary part of $\tau$
\begin{eqnarray}
|| \Phi_{\tau} ||^2 &= \left< \Phi_{\tau} | \Phi_{\tau}\right> = \lim_{D' \rightarrow\infty} \frac{1}{2D'}\int_{-D'}^{D'} \Phi^*_{\tau} (b) \Phi_{\tau} (b) db \nonumber\\
&= \lim_{D'\rightarrow\infty} \frac{1}{2D'}\int_{-D'}^{D'}  |\Phi_0 |^2 \left|\sin\left(\frac{4b}{\hbar}\right)\right| \exp\left(\frac{i\rho_{crit}}{2\hbar} (\tau-\tau^*)\cos\left(\frac{4b}{\hbar}\right) \right)db \nonumber\\
&=  |\Phi_0 |^2 \lim_{D'\rightarrow\infty} \frac{1}{D'}\int_{0}^{D'}  \left|\sin\left(\frac{4b}{\hbar}\right)\right| \exp\left(\frac{-\rho_{crit}}{\hbar} \Im[\tau] \cos\left(\frac{4b}{\hbar}\right) \right) db  \nonumber\\
&=  |\Phi_0 |^2 \lim_{D\rightarrow\infty} \frac{1}{D}\int_{0}^{D}  \left|\sin\left(x\right)\right| \exp\left(s \cos\left(x \right) \right) dx  \nonumber\\
&=  |\Phi_0 |^2 \lim_{ \mathbb{N}\ni n \rightarrow\infty} \frac{1}{2n\pi} \sum_{j=0}^{n-1}\left[ \int_{2j\pi}^{(2j+1)\pi} \left|\sin\left(x\right)\right| \mbox{e}^{s \cos\left(x \right) }dx \right.\nonumber\\
& \;\;\;\;\;\;\;\;\;\;\;\;\;\;\;\;\;\;\;\;\;\;\;\;\;\;\;\;\;\;\;\;\;\;\;\;\;\; \left.
+ \int_{(2j+1)\pi}^{(2j+2)\pi} \left|\sin\left(x\right)\right| \mbox{e}^{s \cos\left(x \right) }dx \right] \nonumber\\
&=  |\Phi_0 |^2 \lim_{ \mathbb{N}\ni n \rightarrow\infty} \frac{1}{2n\pi} \sum_{j=0}^{n-1}\left[ \int_{0}^{\pi} \left|\sin\left(x+2j\pi\right)\right| \mbox{e}^{s \cos\left(x +2j\pi\right) }dx \right.\nonumber\\
& \;\;\;\;\;\;\;\;\;\;\;\;\;\;\;\;\;\;\;\;\;\;\;\;\;\;\;\;\;\;\;\; \left.
+ \int_{0}^{\pi} \left|\sin\left(x+(2j+1)\pi\right)\right| \mbox{e}^{s \cos\left(x + (2j+1)\pi\right) }dx \right] \nonumber\\
&=  |\Phi_0 |^2 \lim_{ \mathbb{N}\ni n \rightarrow\infty} \frac{1}{2n\pi} \sum_{j=0}^{n-1}\left[ \int_{0}^{\pi} \left|\sin\left(x\right)\right| \mbox{e}^{s \cos\left(x \right) }dx \right.\nonumber\\
& \;\;\;\;\;\;\;\;\;\;\;\;\;\;\;\;\;\;\;\;\;\;\;\;\;\;\;\;\;\;\;\;\;\;\;\;\;\; \left.
+ \int_{0}^{\pi} \left|-\sin\left(x\right)\right| \mbox{e}^{-s \cos\left(x\right) }dx \right] \nonumber\\
&=  |\Phi_0 |^2 \lim_{ \mathbb{N}\ni n \rightarrow\infty} \frac{1}{2n\pi} \sum_{j=0}^{n-1} \int_{0}^{\pi} \sin\left(x\right)\left( \mbox{e}^{s \cos\left(x \right) } + \mbox{e}^{-s \cos\left(x \right) } \right) dx \nonumber\\
&=  |\Phi_0 |^2 \lim_{ \mathbb{N}\ni n \rightarrow\infty} \frac{1}{2n\pi} n \int_{0}^{\pi} \sin\left(x\right)2 \cosh\left( s \cos\left(x \right)\right) dx \nonumber\\
&=   \frac{|\Phi_0 |^2}{\pi} \frac{2\sinh(s)}{s} = \frac{2\hbar |\Phi_0 |^2}{\pi} \frac{\sinh\left(\rho_{crit} \Im[\tau]/\hbar\right)}{\rho_{crit} \Im[\tau]} < \infty \nonumber\\
\end{eqnarray}
Where, we again let $x = 4b/\hbar$, $D = 4D'/\hbar$, and $s = -\rho_{crit} \Im[\tau]/\hbar$ in the fourth line. This then shows that even the $\Phi_{\tau}$'s for $\Im[\tau]\neq 0$ are normalizable in $\mathcal{H}^{kin}$. Moreover, including the result for $\tau\in\mathbb{R}$, this in turn shows that all $\Phi_{\tau} \in D(\hat{T}) $ are eigenstates of $\hat{T}$ for any eigenvalue $\tau \in \mathbb{C}$. Indeed, the eigenstates can be normalized with an appropriate choice of $\Phi_0 = (\pi\rho_{crit}\Im[\tau])^{1/2}(2\hbar\sinh(\rho_{crit}\Im[\tau]/\hbar))^{-1/2}$. Note that $\Phi_0 \rightarrow \Phi_0 ^{(\mathbb{R})} $ when $\Im[\tau]\rightarrow 0$.


Thus, $(\tau  \hat{\mathbb{I}} - \hat{T})$ is not invertible for all $\tau \in \mathbb{C}$ so that $\sigma_p(\hat{T}) = \mathbb{C}$ and the continuous and residual spectra are empty. That is, the spectrum of the four-volume operator $\hat{T}$ is purely discrete and is the entire complex plane. This also implies that $\hat{T}$ is non self-adjoint when defined on the maximal domain $D(\hat{T})$ given by Eq (\ref{maxdom}). This does not mean however that $\hat{T}$ has no physical meaning. There are meaningful non self-adjoint quantum operators such as the momentum operator on the half-real line and the time of arrival operators for free non-relativistic and relativistic particles (see, for instance \cite{bu2,bu3,muga}). To study whether $\hat{T}$ is a legitimate Hilbert space operator, one then needs to determine if $\hat{T}$ is maximally symmetric and densely defined. If both are affirmative, then $\hat{T}$ is indeed a legitimate and meaningful Hilbert space operator. This may then be the focus of future studies.

\section{ Possible Self-Adjoint Versions of $\hat{T}$ }
\label{selfadj}
In this section, we consider a family of operators $\hat{T}^{(b_0,\phi_0)}$, each with the same action as $\hat{T}$ given by Eq (\ref{quanT}), and each defined on their respective domains $D(\hat{T}^{(b_0,\phi_0)})\subset \mathcal{H}^{kin}$ given by
\begin{eqnarray}\label{adjdom}
D(\hat{T}^{(b_0,\phi_0)}) = \left\{ \Phi \in \mathcal{H}^{kin}  \left|   ||\hat{T}^{(b_0,\phi_0)}\Phi ||^2   < \infty  \;\; \mbox{and} \;\;  \Phi(b_0) = \mbox{e}^{i\phi_0} \Phi \left(b_0+\pi \frac{\hbar}{4} \right)  \right . \right\} \nonumber\\
\end{eqnarray}
for $\frac{4b_0}{\hbar} \in \mathbb{R} \backslash \{0, \pm\pi/2, \pm\pi, ...\} $ and $\phi_0\in[0,2\pi)$. Thus, $\hat{T}^{(b_0,\phi_0)}$ represents a restriction of $\hat{T}$ with an additional requirement of quasi-periodicity in its domain. It turns out that $\hat{T}^{(b_0,\phi_0)}$ is a possible self-adjoint version of $\hat{T}$. We see this by calculating the discrete spectrum of $\hat{T}^{(b_0,\phi_0)}$. That is, we determine the set of $\tau$'s in which the corresponding formal solutions of $(\tau\hat{\mathbb{I}} - \hat{T}^{(b_0,\phi_0)})\Phi_{\tau} = 0$ are in $D(\hat{T}^{(b_0,\phi_0)})$. Note that the formal solutions $\Phi_{\tau}(b)$ are still given by Eq (\ref{eigT}) since $\hat{T}^{(b_0,\phi_0)}$ and $\hat{T}$ have the same actions. If these $\tau$'s are all real valued, then this suggests that $\hat{T}^{(b_0,\phi_0)}$ is self-adjoint when defined with the domain $D(\hat{T}^{(b_0,\phi_0)})$. However, to explicitly show its self-adjointness, we need to determine whether $\hat{T}^{(b_0,\phi_0)}$ is densely defined and whether $\left< \Phi_a \left| \hat{T}^{(b_0,\phi_0)}\Phi_b \right.\right> = \left< \left.\hat{T}^{(b_0,\phi_0)} \Phi_a \right| \Phi_b \right> $ holds for any $\Phi_a, \Phi_b \in D(\hat{T}^{(b_0,\phi_0)})$. This will then be a topic for future studies. For now however, we will just motivate studying the possible self-adjointness of  $\hat{T}^{(b_0,\phi_0)}$ by showing that all the allowed $\tau$'s are real valued.

Since we have already shown that the $\Phi_{\tau}(b)$'s are normalizable in $\mathcal{H}^{kin}$ for any $\tau\in\mathbb{C}$, then the first condition $||\hat{T}^{(b_0,\phi_0)}\Phi_{\tau} ||^2   < \infty$ is already satisfied. We now turn to impose the condition of quasi-periodicity. Specifically, we first consider the square of its magnitude. From Eq (\ref{eigT}), we calculate
\begin{eqnarray}
\;\;\;\;\;\;\;\;\;\;\;\; |\Phi_{\tau}(b_0)|^2 &=  |\Phi_0 |^2\left|\sin\left(\frac{4b_0}{\hbar}\right)\right| \exp\left(\frac{-\rho_{crit}}{\hbar} \Im[\tau] \cos\left(\frac{4b_0}{\hbar}\right) \right)  \nonumber\\
\left|\Phi_{\tau}\left(b_0 + \pi \frac{\hbar}{4}\right)\right|^2 &=  |\Phi_0 |^2\left|\sin\left(\frac{4b_0}{\hbar} + \pi\right)\right| \exp\left(\frac{-\rho_{crit}}{\hbar} \Im[\tau] \cos\left(\frac{4b_0}{\hbar}+ \pi\right) \right)  \nonumber\\
&= |\Phi_0 |^2\left|\sin\left(\frac{4b_0}{\hbar} \right)\right| \exp\left(\frac{\rho_{crit}}{\hbar} \Im[\tau] \cos\left(\frac{4b_0}{\hbar}\right) \right)  \nonumber\\
\end{eqnarray}
so that $|\Phi_{\tau}(b_0)|^2 = \left|\Phi_{\tau}\left(b_0 + \pi \frac{\hbar}{4}\right)\right|^2$ implies that $\Im[\tau] = 0$. That is, only $\Phi_{\tau}(b)$'s with $\tau\in\mathbb{R}$ can be in $D(\hat{T}^{(b_0,\phi_0)})$. Next, we turn to impose the quasi-periodicity condition itself. 
\begin{eqnarray}
\Phi_{\tau} \left(b_0+\pi \frac{\hbar}{4} \right)  &= \Phi_0 \left|\sin\left(\frac{4b_0}{\hbar}+\pi\right)\right|^{1/2} \exp\left(\frac{i\rho_{crit}\tau}{2\hbar} \cos \left(\frac{4b_0}{\hbar}+\pi\right) \right)\nonumber\\
&= \Phi_0 \left|\sin\left(\frac{4b_0}{\hbar}\right)\right|^{1/2} \exp\left(-\frac{i\rho_{crit}\tau}{2\hbar} \cos \left(\frac{4b_0}{\hbar}\right) \right)\nonumber\\
&= \exp\left(-i \phi_0\right)\Phi_{\tau}(b_0) \nonumber\\
&= \Phi_0 \left|\sin\left(\frac{4b_0}{\hbar}\right)\right|^{1/2} \exp\left(\frac{i\rho_{crit}\tau}{2\hbar} \cos \left(\frac{4b_0}{\hbar}\right) -i \phi_0 \right)\nonumber\\
\nonumber
\end{eqnarray}
which implies that 
\begin{eqnarray}
\exp\left(\frac{i\rho_{crit}\tau}{2\hbar} \cos \left(\frac{4b_0}{\hbar}\right) -i \phi_0 \right) &= \exp\left(-\frac{i\rho_{crit}\tau}{2\hbar} \cos \left(\frac{4b_0}{\hbar}\right) \right)\nonumber\\
\exp\left(\frac{i\rho_{crit}\tau}{\hbar} \cos \left(\frac{4b_0}{\hbar}\right) -i \phi_0 \right)  &= 1\nonumber\\
\;\;\;\;\;\;\;\;\;\;\; \frac{\rho_{crit}\tau_m}{\hbar} \cos \left(\frac{4b_0}{\hbar}\right) - \phi_0 &= 2\pi m\nonumber\\
\;\;\;\;\;\;\;\;\;\;\;\;\;\;\;\;\;\;\;\;\;\;\;\;\;\;\;\;\;\;\;\;\;\;\;\;\;\;\;\;\;\; \tau_m &=  \frac{(\phi_0 + 2\pi m) \hbar}{\rho_{crit} } \sec\left(\frac{4b_0}{\hbar}\right) \label{deigv}
\end{eqnarray}
for $m = 0, \pm 1, \pm 2, ...$. Thus, only the $\Phi_{\tau_m}$'s corresponding to these $\tau_m$'s are in $D(\hat{T}^{(b_0,\phi_0)})$. The discrete spectrum of $\hat{T}^{(b_0,\phi_0)}$ is then the set of $\tau_m$'s given by Eq (\ref{deigv}). Moreover, $\Phi_{\tau_m}$ is then truly an eigenstate with a real eigenvalue $\tau_m$. And since all the eigenvalues are real valued, this then suggests that $\hat{T}^{(b_0,\phi_0)}$ is self-adjoint. As mentioned earlier, however, one still needs to explicitly determine whether $\hat{T}^{(b_0,\phi_0)}$ is truly self-adjoint by proving certain conditions. If so, then the states of definite four-volume $\Phi_{\tau_m}$ do indeed correspond to real and discrete four-volume $\tau_m$. That is, spacetime four-volume is quantized. Also, a choice of $b_0$ and $\phi_0$ is not physically motivated as of the moment. It was suggested that each of the $\hat{T}^{(b_0,\phi_0)}$'s correspond to the restrictions of $\hat{T}$ on a superselection sector determined by the Hamiltonian constraint of the theory such that the $\hat{T}^{(b_0,\phi_0)}$'s are self-adjoint on these superselection sectors. That is, perhaps a choice of a domain $D(\hat{T}^{(b_0,\phi_0)})$ corresponds to a certain superselection sector in which $\hat{T}^{(b_0,\phi_0)}$ is truly self-adjoint. In other words, one may choose a superselection sector in which $\hat{T}^{(b_0,\phi_0)}$ is densely defined on it and satisfies the equality $\left< \Phi_a \left| \hat{T}^{(b_0,\phi_0)}\Phi_b \right.\right> = \left< \left.\hat{T}^{(b_0,\phi_0)} \Phi_a \right| \Phi_b \right> $ for any $\Phi_a, \Phi_b \in D(\hat{T}^{(b_0,\phi_0)})$. This information on the superselection sectors may be encoded in the spectral properties of the operator $\hat{\Lambda}$ since its classical expression $\tilde{\Lambda}$ was derived from the Hamiltonian constraint. Finding the superselection sector in which a choice of $\hat{T}^{(b_0,\phi_0)}$ can be shown to be truly self-adjoint on it will also be the direction of future studies.


%

\section{Conclusions}\label{conc}

In this study, we have constructed a quantum spacetime four-volume operator $\hat{T}$ in the context of ULQC for the $k=0$ FLRW model with no matter sources. Our interpretation of $\hat{T}$ stems from its action on definite volume states $\mathcal{N}_{\nu}(b)$ reducing to the classical four-volume of a finite region in spacetime $\tilde{T}(c,p)$ in the continuum limit $\Delta\rightarrow 0$ and its conjugacy with $\hat{\Lambda}$ corresponding to the classical cosmological constant $\tilde{\Lambda}(c,p)$. The operator $\hat{T}$ was constructed by imposing this conjugacy at the quantum level. The corresponding commutation relation with $\hat{\Lambda}$ was solved by using an expansion of $\hat{T}$ in terms of the Bender-Dunne-like basis operators $\hat{T}_{m,n}$ \cite{art7, art8}. Formal solutions $\Phi_{\tau}$ to the eigenvalue equation $(\tau\hat{\mathbb{I}} - \hat{T})\Phi_{\tau} = 0$ were derived and were shown to be normalizable in the kinematic Hilbert space $\mathcal{H}^{kin}$ for all $\tau\in\mathbb{C}$. Upon assigning the maximal domain $D(\hat{T})$ to $\hat{T}$, we then have $\Phi_{\tau}\in D(\hat{T})$ for all $\tau\in\mathbb{C}$. Thus, the spectrum of $\hat{T}$ is purely discrete and is the entire complex plane. This suggests that $\hat{T}$ is non self-adjoint when defined on the maximal domain $D(\hat{T})$. Investigations in its Hilbert space properties, such as the question of whether $\hat{T}$ is maximally symmetric and densely defined must then be made so that we may clarify its status as a Hilbert space operator. Moreover, the $\Phi_{\tau}$'s are then indeed eigenstates of $\hat{T}$ and can be interpreted as states of definite four-volume $\tau$. It is comfortable to give this interpretation for $\tau\in\mathbb{R}$ but for $\Im[\tau]\neq 0$, an appropriate interpretation is needed and can be addressed in future studies. Also, the operators $\hat{T}^{(b_0,\phi_0)}$, defined on their respective domains $D( \hat{T}^{(b_0,\phi_0)})$, were considered as possible self-adjoint versions of $\hat{T}$. They represent the restrictions of $\hat{T}$ on a domain with certain quasi-periodicity conditions. It was found that their discrete spectra consist of discrete real numbers $\tau_m =  (\phi_0 + 2\pi m) \hbar\rho_{crit}^{-1}  \sec\left(4b_0/\hbar\right)$ for $m = 0,\pm1,\pm2,...$. Since their discrete spectra are all real valued, this then suggests that $\hat{T}^{(b_0,\phi_0)}$ is self-adjoint. However, in order to truly show its self-adjointness, one needs to verify certain conditions. A possible direction here is that perhaps the $D( \hat{T}^{(b_0,\phi_0)})$'s are connected to the superselection sectors determined by the Hamiltonian constraint such that a choice of $D( \hat{T}^{(b_0,\phi_0)})$ may correspond to a certain superselection sector in which $\hat{T}^{(b_0,\phi_0)}$ is self-adjoint on it. That is, there may be a suitable choice such that $\hat{T}^{(b_0,\phi_0)}$ is densely defined on a corresponding superselection sector and satisfies the condition $\left< \Phi_a \left| \hat{T}^{(b_0,\phi_0)}\Phi_b \right.\right> = \left< \left.\hat{T}^{(b_0,\phi_0)} \Phi_a \right| \Phi_b \right> $ for any $\Phi_a, \Phi_b \in D(\hat{T}^{(b_0,\phi_0)})$. This would be the direction of the next study. If there is an appropriate choice in which the four-volume operator can be shown to be self-adjoint, then we indeed see that the spacetime four-volume $\tau_m$ is quantized. This discreteness in the spectrum of $\hat{T}$, and its possible self-adjoint versions $\hat{T}^{(b_0,\phi_0)}$, in LQC is a motivator to construct and study a corresponding quantum spacetime four-volume operator in the full LQG to see whether spacetime is indeed discrete.


Further explorations such as considering the possibility of inverse volume corrections in $\hat{\Lambda}$ can be made. Also, following from \cite{lpap}, it was suggested that perhaps one can get an uncertainty relation from the commutator Eq (\ref{ccrtl}) and get an estimate for the cosmological constant $\Lambda$ which can be tested with the observed value. The possible completeness of the set of eigenstates, the $\Phi_{\tau}$'s, can also be the subject of future works. Since $\hat{T}$ is not unique, perhaps one can construct other operators by considering other appropriate generating functions as defined in \ref{consT}. Or, for a given domain, perhaps one can find a function $ f(\hat{\Lambda})$ so that $\hat{T} + f(\hat{\Lambda})$ is self-adjoint in $\mathcal{H}^{kin}$. In the other direction, an ansatz for $ f(\hat{\Lambda})$ can be made and an appropriate domain may be constructed for such a condition of self-adjointness. Note that the sum is still canonically conjugate to $ \hat{\Lambda}$ for any $f(\hat{\Lambda})$. Extensions to the $k=\pm 1$ FLRW, anisotropic Bianchi, and inhomogeneous models with different matter contents can also be studied. Since singularity resolution in usual LQC is attributed to the discreteness of space, perhaps one can study the implications of four-volume discreteness. These implications include the time evolution of states in ULQC being possibly modified by a discrete time evolution since one cannot have $\frac{\partial\psi}{\partial T} = \lim_{\Delta T \rightarrow 0} \frac{\Delta \psi }{\Delta T}$. This is in parallel with the modifications of the curvature $\mathcal{F}_{ab}^i$ since we cannot express it in terms of holonomies around closed loops of vanishing area but rather those around a finite minimum area $\Delta$. Although one may need to determine the corresponding minimum eigenvalue (if it is indeed discrete) of a four-volume operator $\hat{T}$ in full LQG beforehand since $\Delta$ is the minimum eigenvalue of the area operator in full LQG. Possible modifications of discrete time evolution in effective dynamics, such as singularity resolution, can also be explored. One can also study whether the discreteness survives if one wishes to consider the physical Hilbert space $\mathcal{H}^{phys}$ or the self-dual version $\gamma\rightarrow i$ of the theory. The role of $\hat{T}$ and its implications can also be studied in SFMs and GFTs. That is, instead of the emergence of spacetime as the condensation of quantum "space atoms", would a similar emergence occur if one considers "spacetime atoms"? 

Having constructed a four-volume operator $\hat{T}$ of a finite patch of spacetime, perhaps one can consider other patches (each with their own four-volume operator $\hat{T'}$) with different properties such as non-zero curvatures, isotropies, and matter contents "stiched" together in order to have a many-patch system with each patch corresponding to a "spacetime atom". Perhaps then one can find rules relating the different eigenstates and eigenvalues of the different patches so that the quantum state of the many-patch system can be the quantum superposition of each patch with themselves and each other and so that the smooth continuum of classical spacetime emerges on large scales. This direction of study may be investigated further by following a similar construction in \cite{lattlqc}, where spacetime is put on a lattice with each cell being homogeneous and isotropic. Considering appropriate interactions between nearby cells, a quantum theory was constructed using standard loop quantization techniques.

\section*{Acknowledgements}
The author would like to thank Eric Galapon and Ian Vega for their fruitful comments and discussions which helped this study.

\appendix

\section{The Construction of $\hat{T}$ }\label{consT}
In this appendix, we show the details on the construction of the operator $\hat{T}$. We start by calculating the commutator $\left[\hat{\nu}^k, \hat{\mathcal{N}}_{\ell} \right]$ which we will use to determine the commutator $\left[ \hat{T}_{m,n}, \hat{\mathcal{N}}_{\ell}\right] $ with the Bender-Dunne like basis operators given by Eq (\ref{tmn}). With that, we can impose the canonical commutation relation $\left[ \hat{T}, \hat{\Lambda}\right] = 8\pi i \ell^2_{pl} \hat{\mathbb{I}} $ with $\hat{\Lambda}$ given by Eq (\ref{qL}) so that we can determine a relation for the expansion coefficients of $ \hat{T} = \sum_{m,n}\alpha_{m,n}\hat{T}_{m,n}$. We solve this relation for the $\alpha_{m,n}$'s by constructing an appropriate generating function.

From Eq (\ref{hvccr}), we have $\left[\hat{\nu}, \hat{\mathcal{N}}_{\ell} \right] = \ell\hat{\mathcal{N}}_{\ell}$,  $\hat{\nu}\hat{\mathcal{N}}_{\ell} = \hat{\mathcal{N}}_{\ell} (\hat{\nu} + \ell \hat{\mathbb{I}}) $ and $\hat{\mathcal{N}}_{\ell} \hat{\nu}=  (\hat{\nu} - \ell \hat{\mathbb{I}})\hat{\mathcal{N}}_{\ell} $ so that we can calculate
\begin{eqnarray*}
\left[\hat{\nu}^2,  \hat{\mathcal{N}}_{\ell} \right] & = \hat{\nu} \left[\hat{\nu},  \hat{\mathcal{N}}_{\ell} \right] + \left[\hat{\nu},  \hat{\mathcal{N}}_{\ell} \right]\hat{\nu} = \ell\hat{\nu}  \hat{\mathcal{N}}_{\ell} + \ell \hat{\mathcal{N}}_{\ell} \hat{\nu}\\
&= \ell\hat{\nu}  \hat{\mathcal{N}}_{\ell} + \ell(\hat{\nu} - \ell \hat{\mathbb{I}})\hat{\mathcal{N}}_{\ell} = 2\ell \hat{\nu}  \hat{\mathcal{N}}_{\ell} - \ell^2 \hat{\mathcal{N}}_{\ell} = \hat{\nu}^2\hat{\mathcal{N}}_{\ell} - (\hat{\nu} - \ell \hat{\mathbb{I}})^2\hat{\mathcal{N}}_{\ell}\\
&= \ell \hat{\mathcal{N}}_{\ell} (\hat{\nu} + \ell \hat{\mathbb{I}}) + \ell \hat{\mathcal{N}}_{\ell} \hat{\nu}  = 2\ell \hat{\mathcal{N}}_{\ell} \hat{\nu} + \ell^2 \hat{\mathcal{N}}_{\ell} = \hat{\mathcal{N}}_{\ell} (\hat{\nu} + \ell \hat{\mathbb{I}})^2 - \hat{\mathcal{N}}_{\ell} \hat{\nu}^2 \\ 
\left[\hat{\nu}^3,  \hat{\mathcal{N}}_{\ell} \right] & = \hat{\nu}^2 \left[\hat{\nu},  \hat{\mathcal{N}}_{\ell} \right] + \left[\hat{\nu}^2,  \hat{\mathcal{N}}_{\ell} \right]\hat{\nu} = \ell \hat{\nu}^2 \hat{\mathcal{N}}_{\ell} +  \hat{\nu}^2\hat{\mathcal{N}}_{\ell} \hat{\nu} - (\hat{\nu} - \ell \hat{\mathbb{I}})^2\hat{\mathcal{N}}_{\ell} \hat{\nu} \\
& \;\;\;\;\;\;\;\;= \ell \hat{\nu}^2 \hat{\mathcal{N}}_{\ell}  + \hat{\nu}^2(\hat{\nu} - \ell \hat{\mathbb{I}})\hat{\mathcal{N}}_{\ell} - (\hat{\nu} - \ell \hat{\mathbb{I}})^3\hat{\mathcal{N}}_{\ell}\\
& \;\;\;\;\;\;\;\;=\hat{\nu}^3\hat{\mathcal{N}}_{\ell} - (\hat{\nu} - \ell \hat{\mathbb{I}})^3\hat{\mathcal{N}}_{\ell}\\
&= \left[\hat{\nu},  \hat{\mathcal{N}}_{\ell} \right]\hat{\nu}^2 +\hat{\nu} \left[\hat{\nu}^2,  \hat{\mathcal{N}}_{\ell} \right] = \ell \hat{\mathcal{N}}_{\ell}\hat{\nu}^2 + \hat{\nu} \hat{\mathcal{N}}_{\ell} (\hat{\nu} + \ell \hat{\mathbb{I}})^2 - \hat{\nu}\hat{\mathcal{N}}_{\ell} \hat{\nu}^2\\
&\;\;\;\;\;\;\;\;=\ell \hat{\mathcal{N}}_{\ell}\hat{\nu}^2 + \hat{\mathcal{N}}_{\ell} (\hat{\nu} + \ell \hat{\mathbb{I}})^3 - \hat{\mathcal{N}}_{\ell} (\hat{\nu} + \ell \hat{\mathbb{I}}) \hat{\nu}^2\\
&\;\;\;\;\;\;\;\;= \hat{\mathcal{N}}_{\ell} (\hat{\nu} + \ell \hat{\mathbb{I}})^3 - \hat{\mathcal{N}}_{\ell}\hat{\nu}^3\\
\left[\hat{\nu}^4,  \hat{\mathcal{N}}_{\ell} \right] &
= \hat{\nu}^3 \left[\hat{\nu},  \hat{\mathcal{N}}_{\ell} \right] + \left[\hat{\nu}^3,  \hat{\mathcal{N}}_{\ell} \right]\hat{\nu} = \ell \hat{\nu}^3 \hat{\mathcal{N}}_{\ell} + \hat{\nu}^3\hat{\mathcal{N}}_{\ell}\hat{\nu} - (\hat{\nu} - \ell \hat{\mathbb{I}})^3\hat{\mathcal{N}}_{\ell}\hat{\nu}\\
&\;\;\;\;\;\;\;\;= \ell \hat{\nu}^3 \hat{\mathcal{N}}_{\ell} + \hat{\nu}^3(\hat{\nu} - \ell \hat{\mathbb{I}})\hat{\mathcal{N}}_{\ell} - (\hat{\nu} - \ell \hat{\mathbb{I}})^4\hat{\mathcal{N}}_{\ell} \\
&\;\;\;\;\;\;\;\;=  \hat{\nu}^4 \hat{\mathcal{N}}_{\ell} - (\hat{\nu} - \ell \hat{\mathbb{I}})^4\hat{\mathcal{N}}_{\ell} \\
&= \left[\hat{\nu},  \hat{\mathcal{N}}_{\ell} \right]\hat{\nu}^3 +\hat{\nu} \left[\hat{\nu}^3,  \hat{\mathcal{N}}_{\ell} \right] = \ell \hat{\mathcal{N}}_{\ell}\hat{\nu}^3 + \hat{\nu} \hat{\mathcal{N}}_{\ell} (\hat{\nu} + \ell \hat{\mathbb{I}})^3 - \hat{\nu}\hat{\mathcal{N}}_{\ell} \hat{\nu}^3\\
&\;\;\;\;\;\;\;\;=\ell \hat{\mathcal{N}}_{\ell}\hat{\nu}^3 + \hat{\mathcal{N}}_{\ell} (\hat{\nu} + \ell \hat{\mathbb{I}})^4 - \hat{\mathcal{N}}_{\ell} (\hat{\nu} + \ell \hat{\mathbb{I}}) \hat{\nu}^3\\
&\;\;\;\;\;\;\;\;= \hat{\mathcal{N}}_{\ell} (\hat{\nu} + \ell \hat{\mathbb{I}})^4 -  \hat{\mathcal{N}}_{\ell}\hat{\nu} ^4\\
&\;\;\vdots\\
\left[\hat{\nu}^k,  \hat{\mathcal{N}}_{\ell} \right] &= \hat{\nu}^k \hat{\mathcal{N}}_{\ell} - (\hat{\nu} - \ell \hat{\mathbb{I}})^k\hat{\mathcal{N}}_{\ell} \\
&= \hat{\mathcal{N}}_{\ell} (\hat{\nu} + \ell \hat{\mathbb{I}})^k -  \hat{\mathcal{N}}_{\ell}\hat{\nu} ^k\\
\end{eqnarray*}
which can be proved by induction.
Now from Eq (\ref{tmn}), (note that ${n \choose k}$ terminates the infinite sum into a finite sum so that $k=0,1,...,n$)
\begin{eqnarray}
\left[ \hat{T}_{m,n}, \hat{\mathcal{N}}_{\ell}\right] &= \frac{1}{2^n} \sum_{k=0}^{n}\frac{n!}{k!(n-k)!}\left[ \hat{\nu}^{k}\hat{\mathcal{N}}_{m} \hat{\nu}^{n-k}, \hat{\mathcal{N}}_{\ell}\right] \nonumber \\
&= \frac{1}{2^n} \sum_{k=0}^{n}\frac{n!}{k!(n-k)!} \left[\hat{\nu}^k \hat{\mathcal{N}}_{m},  \hat{\mathcal{N}}_{\ell} \right]\hat{\nu}^{n-k} \nonumber \\ 
&\;\;\;\;\;\;\;\;\; + \frac{1}{2^n} \sum_{k=0}^{n}\frac{n!}{k!(n-k)!} \hat{\nu}^{k}\hat{\mathcal{N}}_{m}\left[\hat{\nu}^{n-k},  \hat{\mathcal{N}}_{\ell} \right]\nonumber \\
&= \frac{1}{2^n} \sum_{k=0}^{n}\frac{n!}{k!(n-k)!} \left[\hat{\nu}^k ,  \hat{\mathcal{N}}_{\ell} \right]\hat{\mathcal{N}}_{m}\hat{\nu}^{n-k} \nonumber \\ 
&\;\;\;\;\;\;\;\;\; + \frac{1}{2^n} \sum_{k'=0}^{n}\frac{n!}{(n-k')!k'!} \hat{\nu}^{n-k'}\hat{\mathcal{N}}_{m}\left[\hat{\nu}^{k'},  \hat{\mathcal{N}}_{\ell} \right] \label{app1} \\ 
&= \frac{1}{2^n} \sum_{k=0}^{n}\frac{n!}{k!(n-k)!} \left(\hat{\nu}^{k} \hat{\mathcal{N}}_{\ell} - \left(\hat{\nu} - \ell \hat{\mathbb{I}}\right)^{k} \hat{\mathcal{N}}_{\ell} \right)\hat{\mathcal{N}}_{m}\hat{\nu}^{n-k} \nonumber \\ 
&\;\;\;\;\;\;\;\;\; + \frac{1}{2^n} \sum_{k=0}^{n}\frac{n!}{(n-k)!k!} \hat{\nu}^{n-k}\hat{\mathcal{N}}_{m} \left( \hat{\mathcal{N}}_{\ell}\left(\hat{\nu} + \ell \hat{\mathbb{I}}\right)^{k} - \hat{\mathcal{N}}_{\ell}\hat{\nu}^{k}  \right) \nonumber \\ 
&= \frac{1}{2^n} \sum_{k=0}^{n}{n \choose k} \hat{\nu}^{k}\hat{\mathcal{N}}_{m+\ell}\hat{\nu}^{n-k} -  \frac{1}{2^n} \sum_{k=0}^{n}{n \choose k} \hat{\nu}^{n-k}\hat{\mathcal{N}}_{m+\ell}\hat{\nu}^{k} \label{app2} \\ 
& + \frac{1}{2^n} \sum_{k=0}^{n}{n \choose k} \hat{\nu}^{n-k}\hat{\mathcal{N}}_{m+\ell}(\hat{\nu} + \ell \hat{\mathbb{I}})^{k} - \frac{1}{2^n} \sum_{k=0}^{n}{n \choose k} (\hat{\nu} - \ell \hat{\mathbb{I}})^{k}\hat{\mathcal{N}}_{m+\ell}\hat{\nu}^{n-k} \nonumber \\
\nonumber 
\end{eqnarray}
where, the second summation of Eq (\ref{app1}) came from the replacement of the dummy index $k=n-k'$ so that $k'=0,1,...,n$ and the first two summations of Eq (\ref{app2}) cancel out because of the same replacement. We can also rewrite the last two summations of Eq (\ref{app2}). Namely,
\begin{eqnarray}
\frac{1}{2^n} \sum_{k=0}^{n}{n \choose k}\hat{\nu}^{n-k}\hat{\mathcal{N}}_{m+\ell}&(\hat{\nu} + \ell \hat{\mathbb{I}})^{k} = \frac{1}{2^n} \sum_{k=0}^{n}{n \choose k}\hat{\nu}^{n-k}\hat{\mathcal{N}}_{m+\ell} \sum_{j=0}^{k} {k \choose j} \ell^j \hat{\nu}^{k-j} \nonumber \\
&= \sum_{k=0}^n\sum_{j=0}^k\frac{\ell^j}{j!2^n}\frac{n!}{(k-j)!(n-k)!}\hat{\nu}^{n-k}\hat{\mathcal{N}}_{m+\ell} \hat{\nu}^{k-j} \nonumber\\
&= \sum_{j=0}^n\sum_{k=j}^{n}\frac{\ell^j}{j!2^n}\frac{n!}{(k-j)!(n-k)!}\hat{\nu}^{n-k}\hat{\mathcal{N}}_{m+\ell} \hat{\nu}^{k-j}\label{app3}\\
&= \sum_{j=0}^{n} \frac{\ell^j}{j!}\frac{1}{2^n} \sum_{k=0}^{n-j} \frac{n!}{k!(n-j-k)!}\hat{\nu}^{n-j-k}\hat{\mathcal{N}}_{m+\ell}\hat{\nu}^k \label{app4} \\
&= \sum_{j=0}^{n} \frac{n!}{j!(n-j)!} \frac{\ell^j}{2^j}\frac{1}{2^{n-j}} \sum_{k=0}^{n-j} \frac{(n-j)!}{k!(n-j-k)!}\hat{\nu}^{n-j-k}\hat{\mathcal{N}}_{m+\ell}\hat{\nu}^k \nonumber\\
&= \sum_{j=0}^{n} {n \choose j}\left(\frac{\ell}{2}\right)^j\hat{T}_{m+\ell,n-j} \nonumber\\
\nonumber
\end{eqnarray}
where, to obtain Eq (\ref{app3}), we note that the double sum with the inner dummy index running as $j=0,1,...,k$ for every $k=0,1,2,...,n$ as the outer dummy index, is equivalent to the double sum with the inner dummy index running as $k = j,j+1,j+2,...,n$ for every $j=0,1,2,...,n$ as the outer dummy index. Also note that, $0\leq j \leq k \leq n$. For Eq (\ref{app4}), we shift indices. Similarly, we can calculate
\begin{eqnarray}
\frac{1}{2^n} \sum_{k=0}^{n}{n \choose k}(\hat{\nu} - \ell \hat{\mathbb{I}})^{k} & \hat{\mathcal{N}}_{m+\ell}\hat{\nu}^{n-k}  = \frac{1}{2^n} \sum_{k=0}^{n}{n \choose k}\sum_{j=0}^{k} {k \choose j} (-\ell)^j \hat{\nu}^{k-j} \hat{\mathcal{N}}_{m+\ell} \hat{\nu}^{n-k} \nonumber \\
&= \sum_{k=0}^{n}\sum_{j=0}^{k}  \frac{(-\ell)^j}{j! 2^n} \frac{n!}{(k-j)!(n-k)!}\hat{\nu}^{k-j}  \hat{\mathcal{N}}_{m+\ell} \hat{\nu}^{n-k} \nonumber\\
&= \sum_{j=0}^{n}\sum_{k=j}^{n}  \frac{(-\ell)^j}{j! 2^n} \frac{n!}{(k-j)!(n-k)!}\hat{\nu}^{k-j}  \hat{\mathcal{N}}_{m+\ell} \hat{\nu}^{n-k} \label{app5} \\
&= \sum_{j=0}^{n} \frac{(-\ell)^j}{j!}\frac{1}{2^n} \sum_{k=0}^{n-j} \frac{n!}{k!(n-j-k)!}\hat{\nu}^k  \hat{\mathcal{N}}_{m+\ell} \hat{\nu}^{n-j-k}\label{app6} \\
&= \sum_{j=0}^{n} \frac{n!}{j!(n-j)!}\frac{(-\ell)^j}{2^j}\frac{1}{2^{n-j}} \sum_{k=0}^{n-j} \frac{(n-j)!}{k!(n-j-k)!}\hat{\nu}^k  \hat{\mathcal{N}}_{m+\ell} \hat{\nu}^{n-j-k}\nonumber \\
&= \sum_{j=0}^{n} {n \choose j}\left(-\frac{\ell}{2}\right)^j\hat{T}_{m+\ell,n-j} \nonumber\\
\nonumber
\end{eqnarray}
with Eq (\ref{app5}) being obtained similarly from Eq (\ref{app3}), and Eq (\ref{app6}) being obtained similarly from Eq (\ref{app4}). We then have
\begin{eqnarray*}
\left[ \hat{T}_{m,n}, \hat{\mathcal{N}}_{\ell}\right] &= \sum_{j=0}^{n} {n \choose j}\left(1-(-1)^j\right) \left(\frac{\ell}{2}\right)^j\hat{T}_{m+\ell,n-j}\\
&= \sum_{j=0}^{\infty} \frac{n!}{(2j+1)!(n-2j-1)!}2\frac{\ell^{2j+1}}{2^{2j+1}} \hat{T}_{m+\ell,n-2j-1}
\end{eqnarray*}
where we just extend the upper limit to infinity for notational purposes since the sum terminates because of ${n \choose 2j+1}$. From Eq (\ref{qL}) and Eq (\ref{asT}), we can then now calculate the commutator
\begin{eqnarray*}
\left[ \hat{T}, \hat{\Lambda} \right] &= -\frac{3}{4\gamma^2 \Delta} \sum_{m,n}\alpha_{m,n} \left(\left[ \hat{T}_{m,n}, \hat{\mathcal{N}}_{4}\right] + \left[ \hat{T}_{m,n}, \hat{\mathcal{N}}_{-4}\right]  \right)\\
&= -\frac{3}{4\gamma^2 \Delta} \sum_{m,n}\alpha_{m,n}  \sum_{j=0}^{\infty}\frac{n! 4^{j+1}\left(\hat{T}_{m+4,n-2j-1} - \hat{T}_{m-4,n-2j-1}\right) }{(2j+1)!(n-2j-1)!}\\
&=\sum_{m,n} \left(  -\frac{3}{4\gamma^2 \Delta} \sum_{j=0}^{\infty}\frac{(n+2j+1)!4^{j+1}}{(2j+1)!n!}\left( \alpha_{m-4,n+2j+1} - \alpha_{m+4,n+2j+1} \right)  \right)  \hat{T}_{m,n} \\
&=  \sum_{m,n} \left(  8\pi i \ell^2_{pl}  \delta_{m,0} \delta_{n,0} \right) \hat{T}_{m,n}
\end{eqnarray*}
so that we have a relation for the $\alpha_{m,n} $'s by equating the coefficients of the $\hat{T}_{m,n}$'s.
This relation can be rewritten in terms of the generating functions
\begin{equation*}
g_{\pm, m,n}(x) = -\frac{3}{4\gamma^2 \Delta} \frac{1}{8\pi i \ell^2_{pl}}\sum_{j=0}^{\infty}\frac{4^{j+1} (n+2j+1)!}{(2j+1)!n!}\alpha_{m\pm 4, n+2j+1} x^{2j+1}
\end{equation*}
where they are vanishing at the origin $g_{\pm, m,n}(0) = 0$, have an odd parity $g_{\pm, m,n}(-x) = -g_{\pm, m,n}(x) $, with second derivatives 
\begin{eqnarray*}
\frac{d^2}{dx^2}g_{\pm, m,n}(x) &= -\frac{3}{4\gamma^2 \Delta} \frac{1}{8\pi i \ell^2_{pl}} \sum_{j=0}^{\infty}\frac{4^{j+2} (n+2j+3)!}{(2j+1)!n!} \alpha_{m\pm 4, n+2j+3} x^{2j+1}\\
&= 4(n+2)(n+1) g_{\pm, m,n+2}(x)
\end{eqnarray*}
and with the $(+)$ and $(-)$ cases related by
\begin{eqnarray*}
g_{-, m+8,n}(x) &= -\frac{3}{4\gamma^2 \Delta} \frac{1}{8\pi i \ell^2_{pl}}\sum_{j=0}^{\infty}\frac{4^{j+1} (n+2j+1)!}{(2j+1)!n!}\alpha_{m+8-4, n+2j+1} x^{2j+1}\\
&= g_{+, m,n}(x)
\end{eqnarray*}
Namely, the relation for the $\alpha_{m,n} $'s can be written as
\begin{eqnarray*}
g_{-, m,n}(1) -  g_{+, m,n}(1) &= \delta_{m,0}\delta_{n,0} \\
g_{-, m,n}(1) -  g_{-, m+8,n}(1) &= \delta_{m,0}\delta_{n,0} 
\end{eqnarray*}
so that the $\alpha_{m,n} $'s are
\begin{equation*}
\alpha_{m\pm 4,n+1} = -8\pi i \ell^2_{pl} \frac{4\gamma^2 \Delta}{3} \frac{1}{4(n+1)} \lim_{x\rightarrow 0} \frac{g_{\pm, m,n}(x) }{x}
\end{equation*}

A very simple generating function that is vanishing at the origin with an odd parity is $g_{-,m,0} = \gamma_m x$ so that 
\begin{eqnarray*}
g_{-,m,2}(x) &=\frac{1}{4(2)} \frac{d^2}{dx^2}g_{-, m,0}(x) = 0 \\
g_{-,m,4}(x) &= \frac{1}{4(4)(3)} \frac{d^2}{dx^2}g_{-, m,2}(x) = 0\\
 &\;\; \vdots
\end{eqnarray*}
That is, $g_{-,m,n = \{2,4,6,...\} }(x) = 0$ and for simplicity, we also let $g_{-,m,1} = 0$ so that $g_{-,m,n =  \{1,3,5,...\}}(x) = 0$ as well. Then, we get
\begin{eqnarray*}
g_{-, m,0}(1) -  g_{-, m+8,0}(1) &= \delta_{m,0}\delta_{0,0} \\
\gamma_m - \gamma_{m+8} &= \delta_{m,0}
\end{eqnarray*}
where we wish to find an expression for $\gamma_m$ by using the recurrence relation for increasing and positive $m$. It can be shown that the operator constructed from such a process is equal to the one constructed from using the recurrence relation for decreasing and negative $m$. Again for simplicity, we let $\gamma_{m = \{0,1,2,3,4,5,6,7\}} = 0$ so that the only non vanishing $\gamma_m$'s are
\begin{eqnarray*}
\gamma_{8} &= - \delta_{0,0} + \gamma_0 = -1 \\
\gamma_{16} &= \gamma_8 = -1 \\
&\;\; \vdots
\end{eqnarray*}
That is, $\gamma_{8j} = -1$ and in turn, $g_{-,8j,0} = - x$ which gives 
\begin{equation*}
\alpha_{8j - 4,0+1} = -8\pi i \ell^2_{pl} \frac{4\gamma^2 \Delta}{3} \frac{1}{4(0+1)} \lim_{x\rightarrow 0} \frac{g_{-, 8j,0}(x) }{x} = i\hbar \kappa\frac{\gamma^2 \Delta}{3}
\end{equation*}
as the only non-vanishing $\alpha_{m,n} $'s for $j=1,2,3,...$ so that from Eq (\ref{asT})
\begin{eqnarray*}
\hat{T} &= \sum_{j=0}^{\infty}\alpha_{8j + 4,1} \hat{T}_{8j + 4,1} = i\hbar \kappa\frac{\gamma^2 \Delta}{3} \sum_{j=0}^{\infty} \frac{1}{2}\left( \hat{\mathcal{N}}_{8j + 4} \hat{\nu} +   \hat{\nu} \hat{\mathcal{N}}_{8j + 4} 
 \right)\\
&=\frac{i\hbar \kappa\gamma^2 \Delta}{6} \left[ \left(\hat{\mathcal{N}}_4\sum_{j=0}^{\infty} \hat{\mathcal{N}}_4^{2j}\right) \hat{\nu} + \hat{\nu}\left(\hat{\mathcal{N}}_4\sum_{j=0}^{\infty} \hat{\mathcal{N}}_4^{2j}\right)  \right] \\
&=\frac{i\hbar \kappa\gamma^2 \Delta}{6} \left[  \hat{\mathcal{N}}_4 \left( \hat{\mathbb{I}} - \hat{\mathcal{N}}_4 ^2 \right)^{-1}\hat{\nu}  + \hat{\nu} \hat{\mathcal{N}}_4 \left( \hat{\mathbb{I}} - \hat{\mathcal{N}}_4 ^2 \right)^{-1} \right] \\
&= \frac{i\hbar \kappa\gamma^2 \Delta}{6} \left[ \left( \hat{\mathcal{N}}_4^{-1} - \hat{\mathcal{N}}_4 \right)^{-1}\hat{\nu}  + \hat{\nu} \left( \hat{\mathcal{N}}_4^{-1} - \hat{\mathcal{N}}_4 \right)^{-1} \right] \\
&= - \frac{i\hbar \kappa\gamma^2 \Delta}{6} \left[ \left( \hat{\mathcal{N}}_4 - \hat{\mathcal{N}}_{-4} \right)^{-1}\hat{\nu}  + \hat{\nu} \left( \hat{\mathcal{N}}_4 - \hat{\mathcal{N}}_{-4} \right)^{-1} \right] 
\end{eqnarray*}
where we used the operations satisfied by the $\hat{\mathcal{N}}_{\lambda} $'s. Other operators $\hat{T}$ may be constructed from other choices of appropriate generating functions $g_{\pm, m,n}(x)$.

\end{document}